# Kernelization Lower Bounds By Cross-Composition[*]

Hans L. Bodlaender[†]    Bart M. P. Jansen[†]    Stefan Kratsch[†]

June 27, 2012

**Abstract**

We introduce the cross-composition framework for proving kernelization lower bounds. A classical problem $L$ AND/OR-cross-composes into a parameterized problem $\mathcal{Q}$ if it is possible to efficiently construct an instance of $\mathcal{Q}$ with polynomially bounded parameter value that expresses the logical AND or OR of a sequence of instances of $L$. Building on work by Bodlaender et al. (ICALP 2008) and using a result by Fortnow and Santhanam (STOC 2008) with a refinement by Dell and van Melkebeek (STOC 2010), we show that if an NP-hard problem OR-cross-composes into a parameterized problem $\mathcal{Q}$ then $\mathcal{Q}$ does not admit a polynomial kernel unless NP $\subseteq$ coNP/poly and the polynomial hierarchy collapses. Similarly, an AND-cross-composition for $\mathcal{Q}$ rules out polynomial kernels for $\mathcal{Q}$ under Bodlaender et al.'s AND-distillation conjecture.

Our technique generalizes and strengthens the recent techniques of using composition algorithms and of transferring the lower bounds via polynomial parameter transformations. We show its applicability by proving kernelization lower bounds for a number of important graphs problems with structural (non-standard) parameterizations, e.g., CLIQUE, CHROMATIC NUMBER, WEIGHTED FEEDBACK VERTEX SET, and WEIGHTED ODD CYCLE TRANSVERSAL do not admit polynomial kernels with respect to the vertex cover number of the input graphs unless the polynomial hierarchy collapses, contrasting the fact that these problems are trivially fixed-parameter tractable for this parameter. We have similar lower bounds for FEEDBACK VERTEX SET and ODD CYCLE TRANSVERSAL under structural parameterizations.

After learning of our results, several teams of authors have successfully applied the cross-composition framework to different parameterized problems. For completeness, our presentation of the framework includes several extensions based on this follow-up work. For example, we show how a relaxed version of OR-cross-compositions may be used to give lower bounds on the degree of the polynomial in the kernel size.

## 1 Introduction

Preprocessing and data reduction are important and widely applied concepts for speeding up polynomial-time algorithms, and for making computation feasible at all in the case of hard problems that are not believed to have efficient algorithms. Kernelization is a way of formalizing data reduction, which allows for a formal analysis of the (im)possibility of data reduction and preprocessing. It originated as a technique to obtain fixed-parameter tractable algorithms for hard (parameterized) problems, and has evolved into its own topic of research (see [32, 3, 44] for recent surveys). A *parameterized problem* [23, 27, 45] is a language $\mathcal{Q} \subseteq \Sigma^* \times \mathbb{N}$, the second component is called the *parameter*. Such a problem is (strongly uniformly) *fixed-parameter tractable* (FPT) if there is a computable function $f$ such that membership of an instance $(x, k)$ can be decided in $f(k)n^{\mathcal{O}(1)}$ time. A *kernelization* algorithm (*kernel*) for $\mathcal{Q}$ transforms an instance $(x, k)$ in time polynomial in $|x| + k$ into an *equivalent* instance $(x', k')$ such that $\max(|x'|, k') \leq f(k)$ for some computable function $f$, which is the *size* of the kernel.

---

[*]This work was supported by the Netherlands Organization for Scientific Research (NWO), project "KERNELS: Combinatorial Analysis of Data Reduction". A preliminary version appeared in the proceedings of the 28th International Symposium on Theoretical Aspects of Computer Science (STACS 2011). Several results have been strengthened compared to the preliminary version.

[†]Department of Information and Computing Sciences, Utrecht University, P.O. Box 80.089, 3508 TB, Utrecht, The Netherlands, {H.L.Bodlaender,B.M.P.Jansen,S.Kratsch}@uu.nl

It is well-known that a decidable problem admits a kernel if and only if it is fixed-parameter tractable (cf. [3]). From a practical perspective we are particularly interested in cases where the kernel size $f$ is polynomially bounded in $k$, so-called *polynomial kernels*. Success stories of kernelization include the $\mathcal{O}(k^2)$ kernel for $k$-VERTEX COVER containing at most $2k$ vertices [12, 13, 1] and the meta-theorems for kernelization of problems on topological graph classes [5, 28]. The research community spent many years trying to find polynomial kernels for elusive problems such as $k$-Path, before techniques were developed which made it possible to prove (under some complexity-theoretic assumption) that a parameterized problem in FPT does not admit a polynomial kernel. Bodlaender et al. [4] introduced the concept of an OR-*composition* algorithm as a tool to give super-polynomial lower bounds on kernel sizes. Informally speaking — we defer formal definitions to Section 2 — an OR-composition algorithm combines a series of inputs of a parameterized problem $\mathcal{Q}$, all sharing the same parameter $k$, into a single instance $(x', k')$ of $\mathcal{Q}$ such that $k'$ is polynomial in $k$, and $(x', k')$ acts as the logical OR of the inputs in the sense that $(x', k')$ is a YES-instance if and only if (at least) one of the inputs is. They used a theorem by Fortnow and Santhanam [29] to show that if there is an OR-composition for an NP-hard parameterized problem $\mathcal{Q}$, then $\mathcal{Q}$ does not admit a polynomial kernelization unless NP $\subseteq$ coNP/poly. This machinery made it possible to prove, e.g., that $k$-PATH and the CLIQUE problem parameterized by the treewidth of the graph do not admit polynomial kernels unless NP $\subseteq$ coNP/poly[1]. The latter condition is equivalent to coNP $\subseteq$ NP/poly, and is deemed unlikely as it implies a collapse of the polynomial hierarchy to its third level [49] and further [11].

It did not take long before the techniques of Bodlaender et al. were combined with the notion of a *polynomial parameter transformation* to also prove lower bounds for problems for which no direct OR-composition algorithm could be found. This idea was used implicitly by Fernau et al. [26] to show that $k$-LEAF OUT-BRANCHING does not admit a polynomial kernel, and was formalized in a paper by Bodlaender et al. [10]: they showed that if there is a polynomial-time transformation from $\mathcal{Q}$ to $\mathcal{Q}'$ which incurs only a polynomial blow-up in the parameter size, then if $\mathcal{Q}$ does not admit a polynomial kernel then $\mathcal{Q}'$ does not admit one either. These polynomial parameter transformations were used extensively by Dom et al. [22] who proved kernelization lower bounds for a multitude of important parameterized problems such as SMALL UNIVERSE HITTING SET and SMALL UNIVERSE SET COVER. Dell and van Melkebeek [21] were able to extend the techniques of Fortnow and Santhanam to prove, e.g., that VERTEX COVER does not admit a kernel of bitsize $\mathcal{O}(k^{2-\varepsilon})$ for any $\varepsilon > 0$. Follow-up work has resulted in lower bounds on the degree of the polynomial in the kernel sizes of packing problems [20, 35], using composition-like techniques in which the output parameter is allowed to depend sublinearly on the number of inputs. The *complementary witness lemma* of Dell and van Melkebeek [21] shows that even co-nondeterministic compositions give kernel lower bounds[2]; this subsequently led one of the current authors to leverage the power of co-nondeterminism to obtain a lower bound for a Ramsey-type problem [39].

**Our contribution.** We introduce the framework of *cross-composition* for obtaining ker-nelization lower bounds, and give several applications to fundamental graph problems under structural parameterizations. The technique generalizes and strengthens the earlier methods of compositions [4] and polynomial-parameter transformations [10], and puts the two existing methods for obtaining kernelization lower bounds in a common perspective. Whereas the existing notion of AND or OR-composition works by composing multiple instances of a *parameterized* problem $\mathcal{Q}$ into a single instance of $\mathcal{Q}$ with a bounded parameter value, in the cross-composition framework it is sufficient to compose the AND or OR of any *classical* NP-hard problem into an instance of the parameterized problem $\mathcal{Q}$ for which we want to prove a lower-bound. The term *cross* in the name stems from this fact: the source- and target problem of the composition need no longer be the same. Since the input to a cross-composition algorithm is a list of *classical* instances instead of parameterized instances, the inputs do not have a parameter in which the output parameter of the composition must be bounded; instead we require that the size of the output parameter is polynomially bounded in the size of the largest input instance. This makes our extension also fruitful when the source and target problems of the composition coincide, since it allows for a larger parameter value in the target instance than what would be permissible in a "classic" composition. In addition we show that the output parameter for a composition of $t$ instances may depend polynomially on the logarithm of the number of input instances, which often simplifies the constructions and proofs. We also introduce the concept of a *polynomial equivalence relation* to remove the need for padding arguments which were frequently required in compositions.

---

[1] In the remainder of this introduction we assume that NP $\not\subseteq$ coNP/poly when stating kernelization lower bounds.

[2] In fact, this is already implicit in the results of Fortnow and Santhanam as observed in unpublished work of Chen and Müller (cf. [33]).

We exhibit the power of cross-composition by giving kernelization lower bounds for *structural* parameterizations of several important graph problems. Since many combinatorial problems are easy on graphs of bounded treewidth [9], and since the treewidth of a graph is bounded by the vertex cover number, many problems are fixed-parameter tractable when parameterized by the cardinality of a given vertex cover. We show that this tractability does not extend to polynomial kernelizability: CLIQUE, CHROMATIC NUMBER, WEIGHTED FEEDBACK VERTEX SET, and WEIGHTED ODD CYCLE TRANSVERSAL do not admit polynomial kernels under this parameterization. In the case of CLIQUE it was already known [4] that the problem does not admit a polynomial kernel parameterized by the treewidth of the graph; since the vertex cover number is at least as large as the treewidth we prove a stronger result. By taking the edge-complement of the input graph, CLIQUE parameterized by the size of a given vertex cover is FPT-equivalent to VERTEX COVER parameterized by the size of a given vertex set whose removal leaves a clique. Hence we also establish a kernel lower bound for the latter problem, which marks an interesting boundary in the search for the smallest parameterization of VERTEX COVER admitting a polynomial kernel [18, 36, 41]. We provide a polynomial-parameter transformation to extend the lower bound to FEEDBACK VERTEX SET and ODD CYCLE TRANSVERSAL parameterized by the vertex-deletion distance to a clique, contrasting the known polynomial kernelizability of these problems with respect to the natural parameterization by solution size [48, 42]. These results strengthen lower bounds given in a preliminary version of this work, where we supplied lower bounds for the smaller parameters "deletion distance to a cluster graph" and "deletion distance to a co-cluster graph".

Since the introduction of the cross-composition framework, it has found numerous applications by various sets of authors [7, 8, 14, 15, 16, 17, 19, 31, 36, 38]. Although all applications of the cross-composition technique presented here are for problems under structural parameterizations, the framework can also be used to obtain lower bounds for natural parameterizations. For example, Cygan et al. [16] employed cross-composition to obtain kernelization lower bounds for EDGE CLIQUE COVER, settling an open problem in kernelization, and also for various graph cut problems parameterized by the size of the cutset. The mentioned follow-up work together with other results on lower bounds for kernelization [21, 20, 35, 39] also suggests several extensions, most of which are reviewed and included in this work. For one, we include the variant of an AND-cross-composition as used first by Cygan et al. [16]. Furthermore, we explicitly define a modification of OR-cross-compositions that allows proofs of polynomial lower bounds for kernelization in the style of Dell and van Melkebeek [21]. Regarding co-nondeterministic variants of OR-cross-compositions we refer the reader to the full version of recent work by Kratsch [39].[3]

**Organization.** We start by giving formal definitions of all concepts relevant for our discussion of kernelization lower bounds in Section 2, thereby providing a brief overview of the existing techniques. In Section 3 we present the cross-composition framework. Our treatise of the subject incorporates several small improvements that have been found after an extended abstract of this work was published [6]. After giving the definitions of the various notions of cross-composition, we show how they imply kernel lower bounds. The remainder of the paper is concerned with applications of the technique to specific parameterized problems. As these are all graph problems, we give some graph-theoretic preliminaries in Section 4.1 before presenting our lower bounds.

## 2 A review of kernelization lower bound techniques

Before introducing the framework of cross-composition, we give formal definitions of the various concepts relating to kernelization. Let us fix some notation. Throughout this work we use $\Sigma$ to denote finite alphabets; note that $\Sigma$ may refer to a different alphabet for each problem. We denote classical problems as $L \subseteq \Sigma^*$ and parameterized problems as $\mathcal{Q} \subseteq \Sigma^* \times \mathbb{N}$. Since tuples $(x, k)$ can straightforwardly be encoded as strings over $\Sigma$ (when $|\Sigma| \geq 2$) we may also say that a parameterized problem is NP-hard or NP-complete; in all cases the parameter value $k$ of $(x, k)$ is encoded in unary, for a total size of $\mathcal{O}(|x| + k)$. We call a string $x \in \Sigma^*$ an *instance for* $L \subseteq \Sigma^*$; ditto for $(x, k) \in \Sigma^* \times \mathbb{N}$ and $\mathcal{Q} \subseteq \Sigma^* \times \mathbb{N}$. We say that the instance $x \in \Sigma^*$ is YES *for* $L \subseteq \Sigma^*$ if $x \in L$, and NO *for* $L$ otherwise; the same holds for $(x, k) \in \mathcal{Q}$. For positive integers $n$ we define $[n] := \{1, \ldots, n\}$.

### 2.1 Kernelizations and Compressions

We begin with a formal definition of kernelization and generalized kernelization (also called *bi-kernel* [2]). Subsequently we also define a natural notion of compressing instances of parameterized problems into short strings over some alphabet, i.e., compressing to some language $L \subseteq \Sigma^*$, which relaxes the notion of kernelization. The known kernelization lower bound techniques, namely via compositions, cross-compositions, and the complementary witness lemma, can all be easily seen to apply also to such compressions. In the following we will state the known tools in this more general way.

**Definition 1** (kernelization, generalized kernelization)**.** Let $\mathcal{Q}, \mathcal{Q}' \subseteq \Sigma^* \times \mathbb{N}$ be parameterized problems and let $h \colon \mathbb{N} \to \mathbb{N}$ be a computable function. A *generalized kernelization for $\mathcal{Q}$ into $\mathcal{Q}'$ of size $h(k)$* is an algorithm that on input $(x, k) \in \Sigma^* \times \mathbb{N}$ takes time polynomial in $|x| + k$ and outputs an instance $(x', k')$ such that:

   **"PB":** The size of $x'$ and the parameter value $k'$ are bounded by $h(k)$.
   
   **"EQ":** The instance $(x', k')$ is YES for $\mathcal{Q}'$ if and only if $(x, k)$ is YES for $\mathcal{Q}$.

The algorithm is a *kernelization*, or in short a *kernel*, for $\mathcal{Q}$ if $\mathcal{Q}' = \mathcal{Q}$. It is a *polynomial (generalized) kernelization* if $h(k)$ is a polynomial.

---
[3]This is not yet available but will soon appear in Transactions in Algorithms.

**Definition 2** (compression)**.** Let $\mathcal{Q} \subseteq \Sigma^* \times \mathbb{N}$ be a parameterized problem, let $L \subseteq \Sigma^*$ be a language, and let $h\colon \mathbb{N} \to \mathbb{N}$ be a computable function. A *compression for $\mathcal{Q}$ into $L$ of size $h(k)$* is an algorithm that on input $(x,k) \in \Sigma^* \times \mathbb{N}$ takes time polynomial in $|x| + k$ and outputs a string $y \in \Sigma^*$ such that:

  **"PB":** The length of $y$ is bounded by $h(k)$.

  **"EQ":** The string $y$ is YES for $L$ if and only if $(x,k)$ is YES for $\mathcal{Q}$.

The algorithm is a *polynomial compression* if $h(k)$ is a polynomial.

The following proposition explicitly states that compressions are a relaxation of generalized kernelizations, which in turn relax the notion of kernelization.

**Proposition 1.** *Let $\mathcal{Q} \subseteq \Sigma^* \times \mathbb{N}$ be a parameterized problem. If $\mathcal{Q}$ has a kernelization with size $h\colon \mathbb{N} \to \mathbb{N}$ then $\mathcal{Q}$ also has a generalized kernelization with size $h$. Similarly, if $\mathcal{Q}$ has a generalized kernelization with size $h$ then $\mathcal{Q}$ has a compression with size $\mathcal{O}(h)$.*

**Remark 1.** The proposition follows immediately from Definitions 1 and 2. Let us clarify however that a compression can be derived from a (generalized) kernelization with kernel size $h$ by mapping each output instance $(x', k') \in \Sigma^* \times \mathbb{N}$ to an encoding as a string $z \in \Sigma^*$ (assume $|\Sigma| \geq 2$, otherwise use $\Sigma' := \{0,1\}$). With a unary encoding for $k'$ this gives a bound of $\mathcal{O}(h(k))$ on the length of $z$ since both $|x'|$ and $k'$ are bounded by $h(k)$, where $k$ is the input parameter.

## 2.2 Compositions and Distillations

We now turn our attention to the main components of the original kernelization lower bound framework of Bodlaender et al. [4]. We will first recall the notions of OR/AND-composition and weak OR/AND-distillations. Both notions formalize ways of encoding the logical OR respectively AND of a large number of instances into a single instance that is small according to some measure. Compositions address parameterized problems and require the output parameter to be small. Distillations address classical problems, or languages, and demand a small overall output size. Although we do not use the compositions of Bodlaender et al. [4] in this work, we supply the definitions to highlight the differences with cross-composition.

**Definition 3** (OR/AND-composition [4])**.** Let $\mathcal{Q} \subseteq \Sigma^* \times \mathbb{N}$ be a parameterized problem. An *OR-composition* for $\mathcal{Q}$ is an algorithm that, given $t$ instances $(x_1, k), \ldots, (x_t, k) \in \Sigma^* \times \mathbb{N}$ of $\mathcal{Q}$, takes time polynomial in $\sum_{i=1}^{t} |x_i| + k$ and outputs an instance $(y, k') \in \Sigma^* \times \mathbb{N}$ such that:

  **"PB":** The parameter value $k'$ is polynomially bounded in $k$.

  **"OR":** The instance $(y, k')$ is YES for $\mathcal{Q}$ if and only if *at least one* instance $(x_i, k)$ is YES for $\mathcal{Q}$.

An *AND-composition* is an algorithm that, instead, fulfills Properties "PB" and "AND".

  **"AND":** The instance $(y, k')$ is YES for $\mathcal{Q}$ if and only if *all* instances $(x_i, k)$ are YES for $\mathcal{Q}$.

A parameterized problem $\mathcal{Q}$ for which such an algorithm exists is called OR-*compositional* respectively AND-*compositional*.

**Definition 4** (weak OR/AND-distillation [4, 29])**.** Let $L, L' \subseteq \Sigma^*$ be languages. A *weak OR-distillation of $L$ into $L'$* is an algorithm that, given $t$ instances $x_1, x_2, \ldots, x_t \in \Sigma^*$ of $L$, takes time polynomial in $\sum_{i=1}^{t} |x_i|$ and outputs a string $y \in \Sigma^*$ such that:

- **"PB":** The length of $y$ is polynomially bounded in $\max_{i=1}^{t} |x_i|$.
- **"OR":** The instance $y$ is YES for $L'$ if and only if *at least one* instance $x_i$ is YES for $L$.

A *weak* AND-*distillation of $L$ into $L'$* is an algorithm that, instead, fulfills Properties "PB" and "AND".

- **"AND":** The instance $y$ is YES for $L'$ if and only if *all* instances $x_i$ are YES for $L$.

The crux of the framework is that the combination of a polynomial kernelization and a composition, for some parameterized problem $\mathcal{Q}$, yields a distillation for $\mathcal{Q}$ (when interpreted as a classical problem). In the following we denote the satisfiability problem for boolean formulae as SAT.

**Theorem 1** ([4]). *Let $\mathcal{Q} \subseteq \Sigma^* \times \mathbb{N}$ be a parameterized problem that is NP-hard when the parameter value is encoded in unary, i.e., when instances $(x,k)$ have size $\mathcal{O}(|x|+k)$. If $\mathcal{Q}$ has a polynomial kernelization (or polynomial compression) and an* OR-*composition, then* SAT *has a weak* OR-*distillation into some language $L \subseteq \Sigma^*$. If $\mathcal{Q}$ instead has an* AND-*composition, then* SAT *has an* AND-*distillation into $L$.*

**Remark 2.** The reason for requiring NP-hardness with parameter value encoded in unary is that this implies a Karp reduction from instances $\phi$ for SAT to instances $(x,k)$ for $\mathcal{Q}$ with $k$ polynomially bounded in $|\phi|$.

Bodlaender et al. [4] conjectured that no NP-hard problem has a distillation.

**Conjecture 1** (OR-distillation conjecture [4]). *No NP-hard problem $L$ has a weak* OR-*distillation into any language $L'$.*

**Conjecture 2** (AND-distillation conjecture [4]). *No NP-hard problem $L$ has a weak* AND-*distillation into any language $L'$.*

The OR-distillation conjecture was proved shortly after by Fortnow and Santhanam [29], under the assumption that $NP \not\subseteq coNP/poly$. Drucker [24] recently announced a proof of the AND-distillation conjecture based on the same assumption, but using much more complicated machinery.

**Theorem 2** ([29]). *If there is a weak* OR-*distillation of* SAT *into any language $L \subseteq \Sigma^*$ then $NP \subseteq coNP/poly$ and the polynomial-time hierarchy collapses to its third level ($PH = \Sigma_3^p$).*

### 2.3 Transformations and Oracle Communication Protocols

The result of Fortnow and Santhanam [29] was generalized by Dell and van Melkebeek [21], who abstract away from the perspective of kernelization and distillation by considering an appropriate oracle communication protocol. They obtain more fine-grained results by taking into account not just the number of instances that are combined but also their maximum size. This way they prove a more general form of Theorem 2, namely their *complementary witness lemma* (see below) and also various polynomial lower bounds for kernelization (and PCPs); e.g., they show that $d$-HITTING SET (i.e., VERTEX COVER on hypergraphs with edge size at most $d$) admits no polynomial kernelization with size $\mathcal{O}(k^{d-\varepsilon})$, for any $\varepsilon > 0$, unless $NP \subseteq coNP/poly$.

**Definition 5** (oracle communication protocol [21]). An *oracle communication protocol* for a language $L$ is a communication protocol for two players. The first player is given the input $x$ and has to run in time polynomial in the length of the input; the second player is computationally unbounded but is not given any part of $x$. At the end of the protocol the first player should be able to decide whether $x \in L$. The cost of the protocol is the number of bits of communication from the first player to the second player.

**Definition 6** (OR/AND of a language [21]). Let $L \subseteq \Sigma^*$ be a language. The OR *of the language* $L$, denoted by $\text{OR}(L)$, is the set of all tuples $(x_1, \ldots, x_t)$ for which $x_i \in L$ for *at least one* $i \in [t]$. The AND *of the language* $L$, denoted by $\text{AND}(L)$, is the set of all tuples $(x_1, \ldots, x_t)$ for which $x_i \in L$ for *all* $i \in [t]$.

**Lemma 1** (complementary witness lemma [21]). *Let $L$ be a language and $t\colon \mathbb{N} \to \mathbb{N}\setminus\{0\}$ be polynomially bounded such that the problem of deciding whether tuples $(x_1, \ldots, x_{t(s)})$ of strings each of length at most $s$ belong to $\text{OR}(L)$ has an oracle communication protocol of cost $\mathcal{O}(t(s)\log t(s))$, where the first player can be co-nondeterministic. Then $L \in \text{coNP/poly}$.*

Finally, we turn our attention to a different way of proving lower bounds for kernelization, namely the use of appropriate reductions. The following type of parameterized reduction, originally called *polynomial time and parameter transformation*, was first formalized by Bodlaender et al. [10]. It is not hard to see that they are strongly related to the existence of polynomial kernelizations (or compressions).

Similarly, one may define parameterized reductions with other bounds on the new parameter value, but they are not required in this work. Reductions with a linear dependence have been used to transfer polynomial bounds for kernelization [35] and in the context of lower bounds on the runtime of FPT-algorithms [43].

**Definition 7** (polynomial parameter transformation [10]). Let $\mathcal{Q}, \mathcal{Q}' \subseteq \Sigma^* \times \mathbb{N}$ be parameterized problems. A *polynomial parameter transformation from $\mathcal{Q}$ to $\mathcal{Q}'$* is an algorithm that on input $(x, k) \in \Sigma^* \times \mathbb{N}$ takes time polynomial in $|x| + k$ and outputs an instance $(x', k') \in \Sigma^* \times \mathbb{N}$ such that:

**"PB":** The parameter value $k'$ is polynomially bounded in $k$.

**"EQ":** The instance $(x', k')$ is YES for $\mathcal{Q}'$ if and only if $(x, k)$ is YES for $\mathcal{Q}$.

We denote the existence of such a transformation by $\mathcal{Q} \leq_{ppt} \mathcal{Q}'$.

**Theorem 3** ([4]). *Let $\mathcal{Q} \subseteq \Sigma^* \times \mathbb{N}$ be a parameterized problem that is NP-hard with parameter encoded in unary, and let $\mathcal{Q}' \subseteq \Sigma^* \times \mathbb{N}$ be a parameterized problem that is contained in NP. If $\mathcal{Q} \leq_{ppt} \mathcal{Q}'$ and $\mathcal{Q}'$ admits a polynomial kernelization then $\mathcal{Q}$ admits a polynomial kernelization.*

To avoid the complication of having to insist on the unary encoding of the parameter value we suggest the following proposition, which applies to the existence of polynomial compressions (and generalized kernelizations). Since lower bounds for kernelization for some parameterized problem $\mathcal{Q} \subseteq \Sigma^* \times \mathbb{N}$ via the tools in this section exclude also polynomial compressions, it seems much easier to only require $\mathcal{Q} \leq_{ppt} \mathcal{Q}'$ to transfer the lower bound to any other parameterized problem $\mathcal{Q}'$.

**Proposition 2.** *Let $\mathcal{Q}$ and $\mathcal{Q}'$ be parameterized problems with $\mathcal{Q} \leq_{ppt} \mathcal{Q}'$. If $\mathcal{Q}'$ admits a polynomial compression then $\mathcal{Q}$ admits a polynomial compression. If, possibly under some complexity-theoretic assumption, $\mathcal{Q}$ admits no polynomial compression then $\mathcal{Q}'$ admits no polynomial compression under the same assumption. The same is true regarding generalized polynomial kernelizations.*

# 3 Cross-composition

## 3.1 The Basic Framework

The two main concepts of the framework are defined as follows.

**Definition 8** (polynomial equivalence relation)**.** An equivalence relation $\mathcal{R}$ on $\Sigma^*$ is called a *polynomial equivalence relation* if the following two conditions hold:

1. There is an algorithm that given two strings $x, y \in \Sigma^*$ decides whether $x$ and $y$ belong to the same equivalence class in time polynomial in $|x| + |y|$.

2. For any finite set $S \subseteq \Sigma^*$ the equivalence relation $\mathcal{R}$ partitions the elements of $S$ into a number of classes that is polynomially bounded in the size of the largest element of $S$.

**Remark 3.** The intended use of polynomial equivalence relations is to group inputs for a (cross-)composition such that the composition will only be applied to groups of instances that are somewhat similar, e.g., they ask for the same solution size or the considered graphs have the same numbers of vertices and edges. We point out that $\mathcal{R} = \Sigma^* \times \Sigma^*$ trivially fulfills the definition of a polynomial equivalence relation (all strings are equivalent and any set $S$ is "partitioned" into a single class). Thus the use of polynomial equivalence relations in cross-compositions according to the following definition is optional.

**Definition 9** (AND/OR-cross-composition)**.** Let $L \subseteq \Sigma^*$ be a language, let $\mathcal{R}$ be a polynomial equivalence relation on $\Sigma^*$, and let $\mathcal{Q} \subseteq \Sigma^* \times \mathbb{N}$ be a parameterized problem. An OR-*cross-composition of $L$ into $\mathcal{Q}$* (with respect to $\mathcal{R}$) is an algorithm that, given $t$ instances $x_1, x_2, \ldots, x_t \in \Sigma^*$ of $L$ belonging to the same equivalence class of $\mathcal{R}$, takes time polynomial in $\sum_{i=1}^{t} |x_i|$ and outputs an instance $(y, k) \in \Sigma^* \times \mathbb{N}$ such that:

**"PB":** The parameter value $k$ is polynomially bounded in $\max_i |x_i| + \log t$.

**"OR":** The instance $(y, k)$ is YES for $\mathcal{Q}$ if and only if *at least one* instance $x_i$ is YES for $L$.

An AND-*cross-composition of $L$ into $\mathcal{Q}$* (with respect to $\mathcal{R}$) is an algorithm that, instead, fulfills Properties "PB" and "AND".

**"AND":** The instance $(y, k)$ is YES for $\mathcal{Q}$ if and only if *all* instances $x_i$ are YES for $L$.

We say that $L$ OR-cross-composes, respectively AND-cross-composes, into $\mathcal{Q}$ if a cross-composition algorithm of the relevant type exists for a suitable relation $\mathcal{R}$.

For historic reasons we reserve cross-composition (without preposition) to refer to the OR variant. As with plain AND/OR-compositions, an AND/OR-cross-composition can be combined with a polynomial compression to give a weak AND/OR-distillation; we prove this explicitly for the case of AND-cross-compositions.

**Theorem 4.** *Let $L \subseteq \Sigma^*$ be a language and let $\mathcal{Q} \subseteq \Sigma^* \times \mathbb{N}$ be a parameterized problem. If $L$ AND-cross-composes into $\mathcal{Q}$ and $\mathcal{Q}$ has a polynomial compression (into an arbitrary language $L'$) then $L$ has a weak AND-distillation (into AND($L'$)).*

*Proof.* To prove the theorem we construct the claimed weak distillation by generalizing the approach of Bodlaender et al. [4]. Let $C$ denote an AND-cross-composition algorithm from $L$ into $\mathcal{Q}$ and let $\mathcal{R}$ denote a suitable polynomial equivalence relation. Let $K$ denote a polynomial compression algorithm for $\mathcal{Q}$ into $L'$. The input to the distillation is a sequence $(x_1, \ldots, x_t)$ of instances of $L \subseteq \Sigma^*$. Define $s := \max_{i=1}^{t} |x_i|$. If $t > (|\Sigma| + 1)^s$ then there must be duplicate inputs, since the number of distinct inputs of length $s' \leq s$ is $|\Sigma|^{s'}$. By discarding duplicates we may therefore assume that $t \leq (|\Sigma| + 1)^s$ and hence that $\log t \in \mathcal{O}(s)$.

The algorithm first partitions the input instances into disjoint nonempty sets $X_1, \ldots, X_r$ according to equivalence under $\mathcal{R}$. By the definition of $\mathcal{R}$ it is clear that this can be done in time polynomial in the input size $\sum_i |x_i|$. Next, the AND-cross-composition algorithm $C$ is applied to each set $X_i$, $1 \leq i \leq r$, and produces one instance $(y_i, k_i) \in \Sigma^* \times \mathbb{N}$ from each set $X_i$. From the definition of an AND-cross-composition and using $\log t \in \mathcal{O}(s)$ it follows that each $k_i$ is polynomially bounded in $s$, and that the computation of these parameterized instances takes time polynomial in the total input size. It also follows that $(y_i, k_i)$ is a YES instance of $\mathcal{Q}$ if and only if all of the instances in $X_i$ are YES for $L$.

Now, we apply the polynomial compression $K$ to each instance $(y_i, k_i)$ to obtain an equivalent instance $z_i$ of $L'$ for each $1 \leq i \leq r$. Since $K$ is a polynomial compression we know that these transformations can be carried out in polynomial time and that $|z_i|$ is polynomially bounded in $k_i$. Since $k_i$ is polynomial in $s$ it follows that $|z_i|$ is also polynomial in $s$ for $1 \leq i \leq r$. Finally, the algorithm simply combines all strings $z_i$ into one tuple $z^* := (z_1, \ldots, z_r)$, which is an instance of AND$(L')$. Since the size of each component is polynomial in $s$, and since the number of components $r$ is polynomial in $s$, we have that $|z^*|$ is polynomial in $s$. By Definition 6 we know that $z^* \in$ AND$(L')$ if and only if all elements of the tuple are contained in $L'$. By tracing back the series of equivalences we therefore find that $z^* \in$ AND$(L')$ if and only if all inputs $x_i$ are YES for $L$. Since we can construct $z^*$ in polynomial time and $|z^*|$ is polynomial in $s$, we have constructed a weak distillation of $L$ into AND$(L')$. □

The analogous result holds for OR-cross-compositions and can be proven along the same lines.

**Theorem 5.** *Let $L \subseteq \Sigma^*$ be a language and let $\mathcal{Q} \subseteq \Sigma^* \times \mathbb{N}$ be a parameterized problem. If $L$ OR-cross-composes into $\mathcal{Q}$ and $\mathcal{Q}$ has a polynomial compression (into an arbitrary language $L'$) then $L$ has a weak OR-distillation (into OR$(L')$).*

The machinery developed in this section leads to two ways of giving kernelization lower bounds, via AND- respectively OR-cross-compositions.

**Corollary 1.** *If an NP-hard language $L$ AND-cross-composes into the parameterized problem $\mathcal{Q}$, then $\mathcal{Q}$ does not admit a (generalized) polynomial kernelization or polynomial compression unless the AND-distillation conjecture fails.*

**Corollary 2.** *If an NP-hard language $L$ OR-cross-composes into the parameterized problem $\mathcal{Q}$, then $\mathcal{Q}$ does not admit a (generalized) polynomial kernelization or polynomial compression unless the OR-distillation conjecture fails and NP $\subseteq$ coNP/poly.*

The lower bounds via OR-cross-compositions are somewhat stronger due to the result of Fortnow and Santhanam (Theorem 2) who proved the OR-distillation conjecture under the assumption that NP $\not\subseteq$ coNP/poly. Note that a weak OR-distillation for an NP-hard language $L$ can be combined with the NP-hardness reduction from SAT to $L$, resulting in a weak OR-distillation for SAT. Using the tight bounds on the compressibility of OR$(L)$ that are provided by the complementary witness lemma, it is possible to derive a statement that can also be applied to give lower bounds on the *degree* of the polynomial for problems that do admit polynomial kernels. We state and prove this extension in the next section as Theorem 6; it also implies the NP $\subseteq$ coNP/poly consequence of Corollary 2 but skips the intermediate step of building a weak OR-distillation.

## 3.2 Cross-compositions of bounded cost

The mentioned recent results of Dell and van Melkebeek [21] that permit polynomial lower bounds via oracle communication protocols for the OR of languages also permit us to create a similar extension for OR-cross-compositions. By a similar analysis as performed by Dell and van Melkebeek [21] (see also [35]) we can establish that appropriate *cross-compositions with bounded cost* imply polynomial lower bounds, assuming NP $\not\subseteq$ coNP/poly.

**Definition 10** (AND/OR-cross-composition of bounded cost)**.** An AND/OR-*cross-composition of $L$ into $\mathcal{Q}$* (with respect to $\mathcal{R}$) *of cost $f(t)$* is an AND/OR-cross-composition algorithm as described in Definition 9 that satisfies "CB" instead of "PB".

> **"CB":** The parameter $k$ is bounded by $\mathcal{O}(f(t) \cdot (\max_i |x_i|)^c)$, where $c$ is some constant independent of $t$.

**Theorem 6.** *Let $L \subseteq \Sigma^*$ be a language, let $\mathcal{Q} \subseteq \Sigma^* \times \mathbb{N}$ be a parameterized problem, and let $d, \epsilon$ be positive reals. If $L$ has an OR-cross-composition into $\mathcal{Q}$ with cost $f(t) = t^{1/d+o(1)}$, where $t$ denotes the number of instances, and $\mathcal{Q}$ has a polynomial compression into an arbitrary language $L'$ with size bound $\mathcal{O}(k^{d-\epsilon})$ then $L \in$ coNP/poly. If, additionally, $L$ is NP-hard then NP $\subseteq$ coNP/poly.*

*Proof.* Let $\mathcal{R}$ be a polynomial equivalence relation on $\Sigma^*$ which partitions any set of strings of length at most $s$ into at most $\mathcal{O}(s^b)$ equivalence classes. Let $f(t) = t^{1/d+o(1)}$ for some constant $d$. Let $C$ be an OR-cross-composition from $L$ into $\mathcal{Q}$, which maps $t$ instances of size at most $s$ and from the same $\mathcal{R}$-equivalence class to an output instance with parameter value bounded by $\mathcal{O}(f(t)s^c)$. Finally, let $K$ be a polynomial compression for $\mathcal{Q}$ into some language $L'$ that outputs a string of size bounded by $h(k) = \mathcal{O}(k^{d-\epsilon})$.

We define a polynomially bounded function $t$ by $t(s) := s^{(b+cd) \cdot \frac{d}{\epsilon}}$. By Lemma 1 it suffices to provide an oracle communication protocol for OR($L$) with cost $\mathcal{O}(t(s) \log t(s))$ for a tuple of $t$ strings each of length at most $s$. Fixing $s$ and $t := t(s)$, let a tuple of $t$ strings each of length at most $s$ be given to the first player, say $(x_1, \ldots, x_t)$.

Let us first describe the protocol and bound the communication cost. The first player begins by partitioning the strings $x_i$ according to equivalence under $\mathcal{R}$, obtaining $r \leq \mathcal{O}(s^b)$ groups. Then he applies the OR-cross-composition $C$ to each group, obtaining $r$ instances $(y_1, k_1), \ldots, (y_r, k_r)$. The parameter values $k_i$ are bounded by $\mathcal{O}(f(t)s^c)$.

Now the first player applies the assumed polynomial compression $K$ to each instance $(y_i, k_i)$, obtaining instances $z_1, \ldots, z_r$ of the language $L'$. Then he sends the instances $z_i$ to the second player, who answers YES if at least one of the instances sent to him is YES, and NO otherwise. (Note that in this simple protocol the second player might as well be an $L'$-oracle, and the cost would correspond to the total length of queries.)

Each compressed instance has size at most $h(k_i) = \mathcal{O}((k_i)^{d-\epsilon}) = \mathcal{O}((f(t)s^c)^{d-\epsilon})$. Thus we can bound the cost of sending the $r$ compressed instances to the second player as follows:

$$\mathcal{O}\left(r \cdot (f(t)s^c)^{d-\epsilon}\right) = \mathcal{O}\left(s^b \left(t^{\frac{1}{d}+o(1)} s^c\right)^{d-\epsilon}\right) = \mathcal{O}\left(s^{b+c(d-\epsilon)} t^{1-\frac{\epsilon}{d}+o(1)}\right) = \mathcal{O}(t),$$

using that $r \leq \mathcal{O}(s^b)$ and the following bound for $s^{b+c(d-\epsilon)}$:

$$s^{b+c(d-\epsilon)} = s^{b+cd} \cdot s^{-c\epsilon} = t^{\frac{\epsilon}{d}} \cdot s^{-c\epsilon} = t^{\frac{\epsilon}{d}-\delta},$$

where $\delta = \frac{c\epsilon^2}{(b+cd)d} > 0$. (Note that $t^{1-\delta+o(1)} = \mathcal{O}(t)$, for any $\delta > 0$.)

**Correctness.** It remains to show correctness of the protocol. Assume first that at least one input instance $x_i$ is a YES-instance (requiring the protocol to correctly answer YES). It follows that the corresponding instance $(y_j, k_j)$ that is created by $C$ from all instances $\mathcal{R}$-equivalent to $x_i$ must be YES for $\mathcal{Q}$. Accordingly, the polynomial compression $K$ transforms $(y_j, k_j)$ to a YES-instance for the language $L'$. Hence, the oracle correctly answers YES, as it is given at least one YES-instance for $L'$.

In the remaining case all input instances $x_1, \ldots, x_t$ are NO for $L$. The OR-cross-composition $C$ will therefore create $r$ NO-instances $(y_i, k_i)$ for $\mathcal{Q}$. These are converted to $r$ NO-instances $z_i$ for $L'$. Hence, presented with only NO-instances for $L'$, the oracle answers NO.

Thus we get an oracle communication protocol for deciding the OR of $t$ instances of $L$ at cost $\mathcal{O}(t)$. By Lemma 1 this implies that $L$ is contained in coNP/poly. Hence, if $L$ is NP-hard then NP $\subseteq$ coNP/poly. □

It is an immediate consequence that cross-compositions with cost $f(t) = t^{o(1)}$ exclude *any* polynomial compression. To see this, observe that their cost is bounded also by $t^{1/d+o(1)}$ for any positive constant $d$. Thus, for any constant $d$ this excludes polynomial compressions to size $\mathcal{O}(k^{d-\epsilon})$ and hence any polynomial compression.

**Corollary 3.** *Let $L \subseteq \Sigma^*$ be a language and let $\mathcal{Q} \subseteq \Sigma^* \times \mathbb{N}$ be a parameterized problem. If $L$ has an OR-cross-composition into $\mathcal{Q}$ with dependence $f(t) = t^{o(1)}$ on the number $t$ of instances and $\mathcal{Q}$ has a polynomial compression into an arbitrary language $L'$ then $L \in$ coNP/poly. If, additionally, $L$ is NP-hard then NP $\subseteq$ coNP/poly.*

## 4 Our lower bounds

Cross-composition makes it possible to give kernelization lower bounds for structural parameterizations of various fundamental graph problems. In the following sections we study CLIQUE, CHROMATIC NUMBER, FEEDBACK VERTEX SET, and ODD CYCLE TRANSVERSAL. We show that parameterizations solvable by the simplest of FPT algorithms nevertheless fail to admit polynomial kernelizations unless NP $\subseteq$ coNP/poly. All lower bounds we give in this section also apply to *generalized* kernelizations, but for brevity we only state them for normal kernelizations. The cross-compositions employed in this section all take the form of OR-cross-compositions. For historic reasons and simplicity we shall therefore omit the adjective OR in the remainder. Before presenting our kernel lower bounds we give some graph-theoretic preliminaries.

### 4.1 Graph preliminaries

We will only work with undirected, finite, simple graphs. For a graph $G$ we denote its vertex set by $V(G)$ and the edge set by $E(G)$. We use $G[V']$ to denote the subgraph of $G$ induced by $V'$, i.e., the graph on vertex set $V'$ and edge set $\{\{u,v\} \in E(G) \mid u,v \in V'\}$. We use $G - Z$ as an abbreviation for $G[V(G) \setminus Z]$. The open neighborhood of a vertex $v$ in graph $G$ contains the vertices adjacent to $v$, and is written as $N_G(v)$. The open neighborhood of a set $S \subseteq V(G)$ is defined as $\bigcup_{v \in S} N_G(v) \setminus S$. *Identifying* a set $S \subseteq V(G)$ of vertices of graph $G$ into a new vertex $x$, results in the graph obtained from $G - S$ by adding a new vertex $x$ adjacent to the vertices in $N_G(S)$. The edge-complement of $G$ is the graph $\overline{G}$ with edge set $\{\{u,v\} \subseteq V(G) \mid u \neq v \wedge \{u,v\} \notin E(G)\}$. An odd cycle is a simple cycle on an odd number of vertices. An $\ell$-coloring of graph $G$ is a function $f: V(G) \to [\ell]$, and it is a *proper* coloring if $\{u,v\} \in E(G)$ implies $f(u) \neq f(v)$.

The chromatic number $\chi(G)$ of the graph $G$ is defined as the smallest integer $\ell$ for which $G$ has a proper $\ell$-coloring. A *vertex cover* in the graph $G$ is a subset $Z \subseteq V(G)$ which intersects all edges, therefore implying that $G - Z$ is an edgeless graph. If $Z$ intersects all cycles then it is a *feedback vertex set* (FVS), and its removal $G - Z$ yields a forest. Finally, an *odd cycle transversal* (OCT) intersects all odd cycles, and its removal results in a bipartite graph. The *vertex cover number* of $G$ is the size of a smallest vertex cover; the numbers related to the other two types of transversals are defined analogously. The vertex-deletion distance of a graph $G$ to a graph family $\mathcal{F}$ is the size of a smallest set $S \subseteq V(G)$ such that $G - S \in \mathcal{F}$. We use $K_t$ and $P_t$ to denote a clique or path on $t$ vertices, respectively. We will occasionally use $n$ to refer to the number of vertices in the (input) graph under consideration. When taking the $n$-bit binary expansion of numbers in the range 1 up to $2^n$, we map $2^n$ to the bitstring consisting of $n$ zeros.

## 4.2 Cliques and Vertex Covers

In this section we prove a kernel lower bound for a strong structural parameterization of CLIQUE, and we consider some of its consequences. An instance of the NP-complete CLIQUE problem [30, GT19] is a tuple $(G, \ell)$ and asks whether the graph $G$ contains a clique on $\ell$ vertices. This classical problem is the source language for the cross-composition which establishes our first kernelization lower bound. The parameterization for which we derive a lower bound is formally defined as follows.

> CLIQUE PARAMETERIZED BY VERTEX COVER
> **Instance:** A graph $G$, a vertex cover $Z \subseteq V(G)$, and a positive integer $\ell$.
> **Parameter:** The size $k := |Z|$ of the vertex cover.
> **Question:** Does $G$ have a clique of size at least $\ell$?

Note that we supply a vertex cover in the input of the problem, to ensure that well-formed instances can be recognized efficiently: the parameter to the problem claims a bound on the vertex cover number of the graph, and using the set $Z$ we may verify this bound in polynomial time. We will define the other parameterized problems for which we establish kernel lower bounds similarly. Observe that these versions of the problems, where a witness of the parameter value is supplied in the input, are certainly no harder to kernelize than the versions where such a witness is not given. An extensive discussion of the technicalities regarding structural parameterizations can be found in the recent survey by Fellows et al. [25].

Let us briefly consider the complexity of the chosen parameterization of the CLIQUE problem. If the graph $G$ has a vertex cover $Z$ and contains a clique $C \subseteq V(G)$, then $|C \setminus Z| \leq 1$: as the definition of a vertex cover ensures there are no edges between vertices in $V(G) \setminus Z$, any clique contains at most one vertex from $V(G) \setminus Z$. The problem defined above can therefore be solved in $\mathcal{O}(2^{|Z|} n^2)$ time by enumerating all subsets $C \subseteq Z$, checking whether they induce a clique $G[C]$, and whether there is a vertex in $V(G) \setminus Z$ adjacent to all members of $C$ which may be added to the clique. We detect a maximum clique when its intersection with $Z$ is used as the subset $C$.

This same insight also yields a so-called *cheating kernel* [26] or *Turing kernel* [34]. For an input instance $(G, Z, \ell, k)$ of CLIQUE PARAMETERIZED BY VERTEX COVER, create a list of instances which contains for each $v \in V(G) \setminus Z$ the instance $(G[Z \cup \{v\}], Z, \ell, k)$, in addition to $(G[Z], Z, \ell, k)$. The preceding discussion shows that $G$ has a clique of size $\ell$, if and only if one of the instances in the list has such a clique. As each instance in the list contains at most $|Z| + 1 = k + 1$ vertices, this is an interesting example of a rare phenomenon: although we cannot, in polynomial time, reduce the size of an instance of CLIQUE PARAMETERIZED BY VERTEX COVER to polynomial in the parameter, we can create $\mathcal{O}(n^{\mathcal{O}(1)})$ instances of size polynomial in the parameter, such that the input is YES if and only if one of the outputs is.

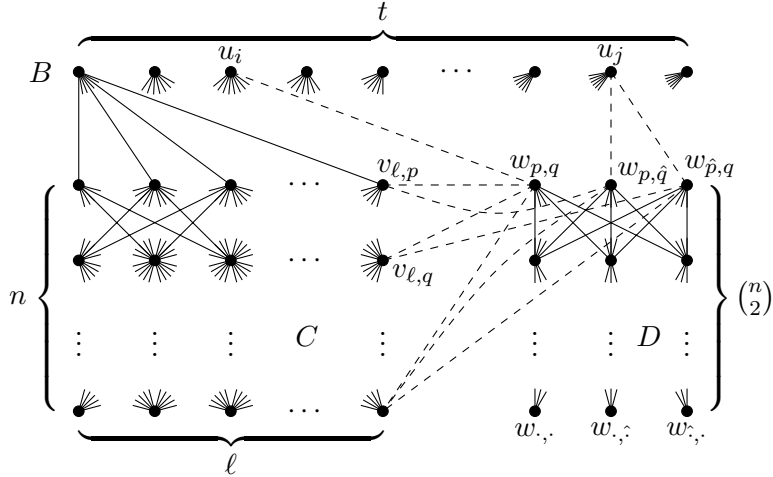

Figure 1: A sketch of the construction used in the proof of Theorem 7. The dashed edges show in an exemplary way how vertices $w_{p,q}$, $w_{p,\hat{q}}$, and $w_{\hat{p},q}$ are connected to vertices of $B$ and $C$, e.g., $\{p,q\}$ is an edge of $G_i$ but not of $G_j$.

Having provided all the necessary background for the following lower bound, we give its proof by cross-composition.

**Theorem 7.** CLIQUE PARAMETERIZED BY VERTEX COVER *does not admit a polynomial kernel unless* $NP \subseteq coNP/poly$.

*Proof.* We prove the theorem by showing that CLIQUE cross-composes into CLIQUE PARAMETERIZED BY VERTEX COVER; by Corollary 2 this is sufficient to establish the claim. We define a polynomial equivalence relation $\mathcal{R}$ such that all bitstrings that do not encode a valid instance of CLIQUE or that have a target clique size $\ell$ that exceeds the number of vertices are equivalent. Of the remaining instances any two (well-formed) instances $(G_1, \ell_1)$ and $(G_2, \ell_2)$ are equivalent if and only if they satisfy $|V(G_1)| = |V(G_2)|$ and $\ell_1 = \ell_2$. From this definition it follows that any set of well-formed instances on at most $n$ vertices each is partitioned into $\mathcal{O}(n^2)$ equivalence classes. Thus $\mathcal{R}$ is a polynomial equivalence relation since the number of bits in a clique instance on $n$ vertices is at least $n$.

We now give a cross-composition algorithm that composes $t$ input instances $x_1, \ldots, x_t$ that are equivalent under $\mathcal{R}$ into a single instance of CLIQUE PARAMETERIZED BY VERTEX COVER. If the input instances are malformed or the size of the clique that is asked for exceeds the number of vertices in the graph, then we may output a single constant-size NO instance; hence in the remainder we may assume that all inputs are well-formed and encode structures $(G_1, \ell), \ldots, (G_t, \ell)$ such that $|V(G_i)| = n$ for all $i \in [t]$ and all instances agree on the value of $\ell \leq n$. We construct a single instance $(G', Z', \ell', k')$ of CLIQUE PARAMETERIZED BY VERTEX COVER, which consists of a graph $G'$ with vertex cover $Z' \subseteq V(G')$ of size $k'$, and an integer $\ell'$.

Let the vertices in each graph $G_i$ be numbered arbitrarily from 1 to $n$. We construct the graph $G'$ as follows (see also Figure 1):

1. Create $\ell n$ vertices $v_{i,j}$ with $i \in [\ell]$ and $j \in [n]$. Connect two vertices $v_{i,j}$ and $v_{i',j'}$ if $i \neq i'$ and $j \neq j'$. Let $C$ denote the set of these vertices. It is crucial that any clique in $G'$ can only contain one vertex $v_{i,\cdot}$ or $v_{\cdot,j}$ for each choice of $i \in [\ell]$ respectively $j \in [n]$. Thus any clique contains at most $\ell$ vertices from $C$.

2. For each pair $1 \leq p < q \leq n$ of distinct vertices from $[n]$ (i.e., vertices of graphs $G_i$), create three vertices: $w_{p,q}$, $w_{p,\hat{q}}$, and $w_{\hat{p},q}$ and make them adjacent to $C$ as follows:

(a) $w_{p,q}$ is adjacent to all vertices from $C$,

(b) $w_{p,\hat{q}}$ is adjacent to all vertices from $C$ except for $v_{\cdot,j}$ with $j = q$, and

(c) $w_{\hat{p},q}$ is adjacent to all vertices from $C$ except for $v_{\cdot,j}$ with $j = p$.

Furthermore we add all edges between vertices $w_{\cdot,\cdot}$ that correspond to distinct pairs from $[n]$. Let $D$ denote these $3\binom{n}{2}$ vertices. Any clique can contain at most one $w_{\cdot,\cdot}$ vertex for each pair from $[n]$.

3. For each input instance $x_i$ with graph $G_i$, make a new vertex $u_i$ and connect it to all vertices in $C$. The adjacency to $D$ is as follows:

   (a) Make $u_i$ adjacent to $w_{p,q}$ if $\{p,q\}$ is an edge in $G_i$.

   (b) Otherwise make $u_i$ adjacent to $w_{p,\hat{q}}$ and $w_{\hat{p},q}$.

   Let $B$ denote this set of $t$ vertices.

We define $\ell' := \ell + 1 + \binom{n}{2}$. Furthermore, we let $Z' := C \cup D$ which is easily verified to be a vertex cover for $G'$ of size $k' := |Z'| = \ell n + 3\binom{n}{2}$. The value $k'$ is the parameter to the problem; it is polynomial in $n$ and hence in the size of the largest input instance. The cross-composition outputs the instance $x' := (G', Z', \ell', k')$. It is easy to see that our construction of $G'$ can be performed in polynomial time. Let us now argue that $x'$ is YES for CLIQUE PARAMETERIZED BY VERTEX COVER if and only if at least one of the instances $x_i$ is YES for CLIQUE.

($\Leftarrow$) First we will assume that some $x_{i^*} = (G_{i^*}, \ell)$ is YES for CLIQUE, i.e., that $G_{i^*}$ contains a clique on at least $\ell$ vertices. Let $S \subseteq [n]$ denote a clique of size exactly $\ell$ in $G_{i^*}$. We will construct a set $S'$ of size $\ell' = \ell + 1 + \binom{n}{2}$ and show that it is a clique in $G'$:

1. We add the vertex $u_{i^*}$ to $S'$.

2. Let $S = \{p_1, \ldots, p_\ell\} \subseteq [n]$. For each $p_j$ in $S$ we add the vertex $v_{j,p_j}$ to $S'$. By Step 1 all these vertices are pairwise adjacent, and by Step 3 they are adjacent to $u_{i^*}$.

3. For each pair $1 \leq p < q \leq n$ there are two cases:

   (a) If $\{p,q\}$ is an edge of $G_{i^*}$ then the vertex $u_{i^*}$ is adjacent to $w_{p,q}$ in $G'$ (by Step 3) and $w_{p,q}$ is adjacent to all vertices of $C$ (by Step 2). We add $w_{p,q}$ to $S'$.

   (b) Otherwise the vertex $u_{i^*}$ is adjacent to both $w_{p,\hat{q}}$ and $w_{\hat{p},q}$. Since the clique $S$ cannot contain both $p$ and $q$ when $\{p,q\}$ is a non-edge, we can add $w_{p,\hat{q}}$ or $w_{\hat{p},q}$ to $S'$ while preserving the fact that $G'[S']$ is a clique. Recall that, e.g., $w_{p,\hat{q}}$ is adjacent to all vertices of $C$ except those corresponding to $q$.

   In both cases we add one $w_{\cdot,\cdot}$-vertex to $S'$, each corresponding to a different pair $p,q$; all these vertices are pairwise adjacent by Step 2.

We have identified the clique $S'$ in $G'$ of size $\ell' = \ell + 1 + \binom{n}{2}$, proving that $x'$ is a YES-instance.

($\Rightarrow$) Now assume that $x'$ is a YES-instance and let $S'$ be a clique of size $\ell + 1 + \binom{n}{2}$ in $G'$. Since $S'$ contains at most $\ell$ vertices from $C$ (i.e., one $v_{i,\cdot}$ for each $i \in [\ell]$) and at most $\binom{n}{2}$ vertices from $D$ it must contain at least one vertex from $B$, say $u_{i^*} \in B$. Since $B$ is an independent set, the clique $S'$ must contain exactly $\ell$ vertices from $C$ and exactly $\binom{n}{2}$ vertices from $D$. Let $S = \{j \in [n] \mid v_{i,j} \in S' \text{ for some } i \in [\ell]\}$. The set $S$ has size $\ell$ since $S'$ contains at most one vertex $v_{\cdot,j}$ for each $j \in [n]$. We will now argue that $S$ is a clique in $G_{i^*}$. Let $p, q \in S$. The clique $S'$ must contain a $w_{\cdot,\cdot}$-vertex corresponding to $\{p,q\}$ and it must contain vertices $v_{i,p}$ and $v_{i',q}$ for some $i, i' \in [\ell]$. Therefore it must contain $w_{p,q}$ since $w_{p,\hat{q}}$ has no edges to vertices $v_{\cdot,q}$ and $w_{\hat{p},q}$ has no edges to $v_{\cdot,p}$ by Step 2. Thus $u_{i^*} \in S'$ must be adjacent to $w_{p,q}$ which implies that $G_{i^*}$ contains the edge $\{p,q\}$. Thus $S$ is a clique in $G_{i^*}$.

Since we proved that the instance $(G', Z', \ell', k')$ can be constructed in polynomial time and that it acts as the OR of the input instances, and because the parameter value $k'$ is bounded by a polynomial in the size of the largest input instance, this concludes the cross-composition proof and establishes the claim. □

In the remainder of this work we will not be as explicit in giving the definition of the equivalence relationship $\mathcal{R}$ as we were in the proof of Theorem 7: we shall simply state the desired forms of the inputs when it is easy to see that such a form can be achieved by a suitable choice of $\mathcal{R}$. We restrict ourselves to giving cross-compositions for well-formed input instances, since we can simply output a constant-sized NO instance if the input is a sequence of malformed instances. The previous theorem yields an interesting corollary for the following structural parameterization of VERTEX COVER.

> VERTEX COVER PARAMETERIZED BY CLIQUE DELETION SET
> **Instance:** A graph $G$, a set $Z \subseteq V(G)$ such that $G - Z$ is a clique, and a positive integer $\ell$.
> **Parameter:** The size $k := |Z|$ of the deletion set.
> **Question:** Does $G$ have a vertex cover of size at most $\ell$?

As the vertex cover number of a clique $K_t$ is $t-1$, we can interpret this as a parameterization by the distance from trivially solvable instances as discussed by Niedermeier [46]. Our results imply that even though the parameter will often be very large, the problem is unlikely to admit a polynomial kernel.

**Corollary 4.** *If $\mathcal{F}$ is a class of graphs containing all cliques, then* VERTEX COVER *and* INDEPENDENT SET *parameterized by the minimum number of vertex deletions to obtain a graph in $\mathcal{F}$ do not admit polynomial kernels unless $NP \subseteq coNP/poly$. In particular,* VERTEX COVER *and* INDEPENDENT SET *parameterized by clique deletion set do not admit polynomial kernels unless $NP \subseteq coNP/poly$.* □

*Proof.* Consider an instance $(G, Z, \ell, k)$ of CLIQUE PARAMETERIZED BY VERTEX COVER. Since a clique in $G$ is an independent set in $\overline{G}$, the CLIQUE instance is equivalent to asking whether the graph $\overline{G}$ has an independent set of size at least $\ell$. Because $Z$ is a vertex cover for $G$ we know that $G - Z$ is an independent set, and therefore $\overline{G} - Z$ is a clique. Hence the instance $(G, Z, \ell, k)$ of CLIQUE PARAMETERIZED BY VERTEX COVER is equivalent to an instance $(\overline{G}, Z, \ell, k)$ of INDEPENDENT SET PARAMETERIZED BY CLIQUE DELETION SET. Since $\overline{G}$ has an independent set of size $\ell$ if and only if it has a vertex cover of size $|V(G)| - \ell$ it follows that these two instances are also equivalent to the instance $(\overline{G}, Z, |V(G)| - \ell, k)$ of VERTEX COVER PARAMETERIZED BY CLIQUE DELETION SET. Hence the described operations serve as polynomial parameter transformations from CLIQUE PARAMETERIZED BY VERTEX COVER into the mentioned problems, which yields the corollary by Theorem 7 and Proposition 2. □

### 4.3 Chromatic Number

In this section we give a kernelization lower bound for CHROMATIC NUMBER PARAMETERIZED BY VERTEX COVER, by cross-composing a restricted version of 3-COLORING into it. We consider the problem on the following class of graphs.

**Definition 11.** A graph $G$ is a *triangle split graph* if $V(G)$ can be partitioned into sets $X$ and $Y$ such that $G[X]$ is an edgeless graph and $G[Y]$ is a disjoint union of triangles.

We will see that 3-Coloring instances on triangle split graphs lend themselves better to composition than general instances: if we take the disjoint union of a set of triangle split graphs and merge the induced triangles one by one, no adjacency information is lost. Before we give the composition, however, we must prove that 3-Coloring is still NP-complete with this restriction on the input graphs.

> 3-Coloring with Triangle Split Decomposition
> **Instance:** A graph $G$ with a partition of its vertex set into $X \cup Y$ such that $G[X]$ is edgeless and $G[Y]$ is a union of vertex-disjoint triangles.
> **Question:** Is there a proper 3-coloring of $G$?

We prove the NP-completeness of this problem by replacing edges in a general instance of 3-Coloring by triangles.

**Lemma 2.** *3-Coloring with Triangle Split Decomposition is NP-complete.*

*Proof.* It is well-known that 3-Coloring on general graphs is NP-complete [30, GT4], and it is trivial to see that the problem restricted to triangle split graphs is contained in NP. We show how to transform an instance $G$ of 3-coloring in polynomial time into an equivalent instance of 3-coloring on a graph $G'$ with a triangle split decomposition of $V(G')$ into sets $X'$ and $Y'$. Number the edges in $G$ as $e_1, e_2, \ldots, e_m$. Construct the graph $G'$ as follows:

- Set $V(G') := V(G) \cup \{a_i, b_i, c_i \mid i \in [m]\}$ and $E(G') := \emptyset$.

- Add the edges $\{a_i, b_i\}, \{b_i, c_i\}, \{a_i, c_i\}$ to $E(G')$ for $i \in [m]$.

- For each edge $e_i = \{u_i, v_i\}$ ($i \in [m]$) of graph $G$, make vertex $u_i$ adjacent in $G'$ to $a_i$, and make $v_i$ adjacent to $b_i$ and $c_i$.

- Define $X' := V(G)$ and $Y' := \{a_i, b_i, c_i \mid i \in [m]\}$.

This concludes the description of $G'$. It is easy to see that $G'$ is a triangle split graph with the partition $X'$ and $Y'$ since $G'[X']$ is an independent set and $G'[Y']$ is a disjoint union of triangles. We now show that $\chi(G') \leq 3$ if and only if $\chi(G) \leq 3$.

($\Rightarrow$) Assume that $\chi(G') \leq 3$ and consider a 3-coloring of $G'$. For every edge $\{u_i, v_i\} \in E(G)$ we added a triangle on vertices $\{a_i, b_i, c_i\}$ to the graph $G'$. Hence $G'[N_{G'}(\{u_i, v_i\})]$ contains a triangle for all pairs of vertices $\{u_i, v_i\}$ which are adjacent in $G$. If some pair $u_i$ and $v_i$ receive the same color, then this leaves only two colors to use on the triangle in their open neighborhood. As it takes three colors to properly color a triangle, any proper 3-coloring of $G'$ uses different colors for $u_i$ and $v_i$, for all $\{u_i, v_i\} \in E(G)$. Therefore the 3-coloring of $G'$ restricted to the vertex set of $G$ is a proper 3-coloring of $G$.

($\Leftarrow$) Assume that $G$ has a proper 3-coloring. We construct a 3-coloring for $G'$ by coloring all vertices of $V(G') \cap V(G)$ the same as in $G$; now all that remains is to color the triangles we added to the graph. If there is a triangle $\{a_i, b_i, c_i\}$ for a pair $\{u_i, v_i\}$ then $\{u_i, v_i\}$ are adjacent in $G$ and hence they receive different colors in the proper coloring. Now give $a_i$ the color of $v_i$, give $b_i$ the color of $u_i$ and give $c_i$ the remaining color. If we do this for every triangle then we obtain a proper 3-coloring of $G'$ which proves that $\chi(G') \leq 3$.

Since the instance $(G', X', Y')$ can be built from $G$ in polynomial time this proves that 3-Coloring with Triangle Split Decomposition is NP-complete. □

Equipped with this lemma we can prove a kernel lower bound for the following problem.

CHROMATIC NUMBER parameterized by VERTEX COVER
**Instance:** A graph $G$, a vertex cover $Z \subseteq V(G)$, and a positive integer $\ell$.
**Parameter:** The size $k := |Z|$ of the vertex cover.
**Question:** Is there a proper $\ell$-coloring of $G$?

The parameterization is easily seen to be fixed-parameter tractable by noting that the chromatic number of a graph exceeds its vertex cover number by at most one. Hence $\ell \leq |Z| + 1$ for all relevant inputs, which yields an FPT algorithm by trying all $\ell$-colorings of $G[Z]$ and testing whether they can be extended to the entire graph. This test comes down to checking whether for each vertex $v \in V(G) \setminus Z$ there is a color which is not yet used on a neighbor in $Z$, resulting in a total runtime of $\mathcal{O}(\ell^{|Z|} n^{\mathcal{O}(1)}) = \mathcal{O}((k+1)^k n^{\mathcal{O}(1)})$.

**Theorem 8.** CHROMATIC NUMBER parameterized by VERTEX COVER *does not admit a polynomial kernel unless* $NP \subseteq coNP/poly$.

*Proof.* To prove the theorem we will show that 3-COLORING WITH TRIANGLE SPLIT DECOMPOSITION cross-composes into CHROMATIC NUMBER parameterized by VERTEX COVER. By a suitable choice of polynomial equivalence relation — in the same style as in Theorem 7 — we may assume that we are given $t$ input instances that encode structures $(G_1, X_1, Y_1), \ldots, (G_t, X_t, Y_t)$ of 3-COLORING WITH TRIANGLE SPLIT DECOMPOSITION with $|X_i| = n$ and $|Y_i| = 3m$ for all $i \in [t]$ (i.e., $m$ is the number of triangles in each instance). We will compose these instances into one instance $(G', Z', \ell', k')$ of CHROMATIC NUMBER parameterized by VERTEX COVER. By duplicating some instances we may assume that the number of inputs $t$ is a power of 2; this does not affect the truth value of the OR of the inputs, and a parameter value which is suitably bounded with respect to the longer sequence of inputs, is also suitably bounded with respect to the original list of inputs.

For each set $Y_i$, label the triangles in $G_i[Y_i]$ as $T_1, \ldots, T_m$ in some arbitrary way, and label the vertices in each triangle $T_j$ for a set $Y_i$ as $a_j^i, b_j^i, c_j^i$. We build a graph $G'$ with a vertex cover of size $k' := 3 \log t + 4 + 3m \in \mathcal{O}(m + \log t)$ such that $G'$ can be $\ell' := \log t + 4$-colored if and only if one of the input instances can be 3-colored.

1. Initialize the graph $G'$ as the disjoint union of the input graphs $G_1, \ldots, G_t$.

2. For each $j \in [m]$, identify the vertices $\{a_j^i \mid i \in [t]\}$ into a new vertex $a_j$. Similarly identify all $b_j^i$ into $b_j$, and all $c_j^i$ into $c_j$, for each $j \in [m]$. The vertices resulting from these identification operations are the *triangle vertices* $T'$. As we are effectively merging triangles one by one, $G'[T']$ is a disjoint union of triangles.

3. Add a clique on vertices $\{p_i \mid i \in [\log t]\} \cup \{w, x, y, z\}$ to $G'$; it is called the *palette*.

4. Make all vertices in $T'$ adjacent to all vertices from the palette except $x, y$, and $z$.

5. Make all the vertices in $\bigcup_{i=1}^{t} X_i$ adjacent to $w$.

6. For $i \in [\log t]$ add a path on two new vertices $\{q_0^i, q_1^i\}$ to the graph, and make them adjacent to all vertices of the palette except $p_i$ and $w$. These vertices form the *instance selector* vertices.

7. For each instance number $i \in [t]$ we can write a binary representation of the value $i$ in $\log t$ bits. Consider each position $j \in [\log t]$ of this binary representation, where position 1 is most significant and $\log t$ is least significant. If bit number $j$ of the representation of $i$ is a 0 (resp. a 1) then make vertex $q_0^j$ (resp. $q_1^j$) adjacent to all vertices of $X_i$.

This concludes the construction. The following claims about $G'$ are easy to verify:

(I) In every proper $\ell' = \log t + 4$-coloring of $G'$, the following holds:

(a) each of the $\log t + 4$ vertices of the palette clique receives a unique color,

(b) for each $i \in [\log t]$ the vertices $q_0^i$ and $q_1^i$ receive different colors (since they are adjacent); one of them must take the color of $w$ and the other of $p_i$ (they are adjacent to all other vertices of the palette),

(c) the triangle vertices $T'$ are colored using the colors of $x, y, z$ (they are adjacent to all other vertices of the palette),

(d) the only colors which can occur on a vertex in $X_i$ (for all $i \in [t]$) are the colors given to $x, y, z$ and $\{p_j \mid j \in [\log t]\}$ (since the vertices in $X_i$ are adjacent to $w$).

(II) For every $i \in [t]$, the graph $G'[X_i \cup T']$ is isomorphic to $G_i$.

(III) The set $Z' := \{p_i \mid i \in [\log t]\} \cup \{w, x, y, z\} \cup T' \cup \{q_0^i, q_1^i \mid i \in [\log t]\}$ forms a vertex cover of $G'$ of size $k' = |Z'| = 3\log t + 4 + 3m$. Hence we establish that $G'$ has a vertex cover of size $\mathcal{O}(m + \log t)$.

The output of the cross-composition is the instance $(G', Z', \ell' := \log t + 4, k')$. It is easy to verify that the construction can be performed in polynomial time. As the value of $k'$ is polynomial in $\log t$ plus the size of the largest input, it remains to prove that the output indeed acts as the logical OR of the input instances: $\chi(G') \leq \log t + 4$ if and only if there is an $i \in [t]$ such that $\chi(G_i) \leq 3$.

($\Rightarrow$) Suppose $\chi(G') \leq \ell'$ and consider some proper $\ell'$-coloring of $G'$. By (I.b) we know that for each $i \in [\log t]$ exactly one vertex of the pair $\{q_0^i, q_1^i\}$ receives the same color as $p_i$. Consider the string of $\log t$ bits where the $i$-th most significant bit is a 1 if and only if vertex $q_1^i$ receives the same color as $p_i$. This bitstring encodes some integer $i^* \in [t]$. We focus on the instance with the number $i^*$. Let $Q$ be the set of vertices which contains for each pair $\{q_0^i, q_1^i\}$ ($i \in [\log t]$) the unique vertex which is colored the same as $p_i$. By the definition of $G'$ we know that all vertices of $X_{i^*}$ are adjacent to all vertices of $Q$; hence in any proper coloring of $G'$ the vertices of $X_{i^*}$ cannot use any colors which are used on $\{p_i \mid i \in [\log t]\}$. By (I.d) this implies that the coloring for $G'$ can only use the colors of $x, y, z$ on the vertices of $X_{i^*}$. By (I.c) the triangle vertices $T'$ are also colored using only the colors of $x, y, z$. The graph $G'[X_{i^*} \cup T']$ is isomorphic to the input graph $G_{i^*}$ by (II), and since the coloring of $G'$ only uses the colors of $x, y, z$ on these vertices, this shows that the coloring of $G'$ restricted to the induced subgraph $G'[X_{i^*} \cup T']$ is in fact a 3-coloring of graph $G_{i^*}$, which proves that $\chi(G_{i^*}) \leq 3$ and establishes this direction of the equivalence.

($\Leftarrow$) Suppose $\chi(G_{i^*}) \leq 3$ for some $i^* \in [t]$. We will construct a proper $\ell'$-coloring of $G'$. Start by giving all vertices of the palette different colors. By (II) the graph $G'[X_{i^*} \cup T']$ is isomorphic to $G_{i^*}$. Re-label the colors in the 3-coloring of $G_{i^*}$ such that it uses the colors given to $\{x, y, z\}$ in our partial $\ell'$-coloring of $G'$. Give a vertex $v$ in the induced subgraph $G'[X_{i^*} \cup T']$ the same color as the vertex in $G_{i^*}$ that it is mapped to by the isomorphism. Afterwards we have a proper partial $\ell'$-coloring, where all vertices of the palette, all vertices of $X_{i^*}$, and all triangle vertices of $G'$ are colored. It remains to color the sets $X_i$ for $i \neq i^*$, and the pairs $\{q_0^i, q_1^i\}$. For each $i \in [\log t]$ we color the pair $\{q_0^i, q_1^i\}$ as follows: if the $i$-th most significant bit of the binary representation of the number $i^*$ is a 1 then we color $q_1^i$ the same color as $p_i$ and we color $q_0^i$ as $w$; if the bit is a 0 then we do it the other way around. It is straight forward to verify that we do not create any monochromatic edges in this way. As the final step we have to color the sets $X_i$ for $i \neq i^*$; so consider some $i \in [t]$ with $i \neq i^*$. The binary representation of the number $i^*$ must differ from the binary representation of $i$ in at least one position; suppose they differ at position $j$. The vertex of $\{q_0^j, q_1^j\}$ that matches the bit value of $i^*$ at position $j$ was colored the same as $p_j$, hence the other vertex of the pair must have been colored the same as $w$. Since the bit values differ, by the definition of adjacencies in $G'$ we find that the vertices $X_i$ are adjacent to the vertex of $\{q_0^j, q_1^j\}$ that is colored as $w$. Therefore the vertices of $X_i$ do not have any neighbors colored as $p_j$, and since $X_i$ is an independent set we may color all vertices in it the same as $p_j$. If we color all sets $X_i$ for $i \neq i^*$ in this way we obtain a proper $\ell'$-coloring of $G'$ which proves that $\chi(G') \leq \ell'$. □

Theorem 8 shows that CHROMATIC NUMBER PARAMETERIZED BY VERTEX COVER is unlikely to admit a polynomial kernel. If we treat the number of colors as a constant $q$, rather than a variable, then the problem *does* admit a polynomial kernel: $q$-COLORING PARAMETERIZED BY VERTEX COVER has a kernel with $\mathcal{O}(k^q)$ vertices for every fixed $q$ [37]. The lower bound shows that $q$ must appear in the degree of the polynomial for such kernels.

By interpreting CHROMATIC NUMBER on $G$ as the problem of partitioning the vertex set of $\overline{G}$ into a minimum number of cliques, we obtain the following corollary to Theorem 8 — note that if $Z \subseteq V(G)$ is a vertex cover of $G$, then $\overline{G} - Z$ is a clique.

**Corollary 5.** VERTEX PARTITION INTO CLIQUES *[30, GT15] parameterized by vertex-deletion distance to a clique does not admit a polynomial kernel unless* $NP \subseteq coNP/poly$.

## 4.4 Feedback Vertex Set and Odd Cycle Transversal

The problems FEEDBACK VERTEX SET (FVS) and ODD CYCLE TRANSVERSAL (OCT) ask whether a given graph contains a small set of vertices that intersects all cycles, respectively all odd cycles. The study of their natural parameterization by the solution size has led to important advances in the design of FPT algorithms and kernelizations. Both problems are fixed-parameter tractable [48, 47] and admit (randomized) polynomial kernels [48, 42]. We consider the kernelization complexity of these problems from a different angle. First we study the non-standard parameter "deletion distance to a clique" and use a polynomial parameter transformation to transfer a lower bound result from the similarly parameterized vertex cover problem (Corollary 4). Thus the two cycle transversal problems do not admit polynomial kernels for this parameter, unless NP $\subseteq$ coNP/poly. Then we consider a generalization of the problem where each vertex is assigned an integral weight, and we seek a cycle transversal of bounded total weight. Using cross-composition we prove that the resulting problems do not admit polynomial kernels when parameterized by the cardinality of a given vertex cover.

### 4.4.1 Unweighted problems parameterized by distance to a clique

We give kernelization lower bounds for FVS and OCT under the following structural parameterization.

> FEEDBACK VERTEX SET PARAMETERIZED BY CLIQUE DELETION SET
> **Instance:** A graph $G$, a set $Z \subseteq V(G)$ such that $G - Z$ is a clique, and a positive integer $\ell$.
> **Parameter:** The size $k := |Z|$ of the deletion set.
> **Question:** Is there a set $S \subseteq V(G)$ of at most $\ell$ vertices that intersects all cycles?

The problem ODD CYCLE TRANSVERSAL PARAMETERIZED BY CLIQUE DELETION SET is defined analogously, by asking for a set that intersects all *odd* cycles. It is easy to give FPT algorithms for these two parameterizations. If $G$ is a graph in which the removal of $Z \subseteq V(G)$ leaves a clique, then an FVS or OCT in $G$ avoids at most two vertices of $V(G) \setminus Z$, as otherwise $V(G) \setminus Z$ contains a triangle not intersected by the transversal. Hence there are $\mathcal{O}(n^2)$ ways in which a valid transversal may intersect $V(G) \setminus Z$. For each valid intersection with $V(G) \setminus Z$ we may simply try all subsets of $Z$ to see which gives the smallest transversal that is valid for the entire graph, resulting in an algorithm of runtime $\mathcal{O}(2^{|Z|} n^{\mathcal{O}(1)})$.

Despite the simplicity of the algorithms, the problems are unlikely to admit polynomial kernels. The following theorem strengthens the results in a preliminary version of this work where we proved kernel lower bounds for FVS under the smaller parameterizations "vertex-deletion distance to a cluster graph" and "vertex-deletion distance to a co-cluster graph" [6]. The theorem also subsumes results from an extended abstract by a subset of the authors, where a kernel lower bound for OCT was given for parameterizations by (co-)cluster deletion distance [38].

**Theorem 9.** FEEDBACK VERTEX SET *and* ODD CYCLE TRANSVERSAL *do not admit polynomial kernels when parameterized by a clique deletion set, unless* $NP \subseteq coNP/poly$.

*Proof.* By Corollary 4 and Proposition 2 it is sufficient to prove that there are a polynomial-parameter transformations from VERTEX COVER PARAMETERIZED BY CLIQUE DELETION SET into FVS PARAMETERIZED BY CLIQUE DELETION SET and OCT PARAMETERIZED BY CLIQUE DELETION SET. We give a single construction that works for both cases.

Let $(G, Z, \ell, k)$ be an instance of VERTEX COVER PARAMETERIZED BY CLIQUE DELETION SET. Let us first argue that the edges of $G$ can be covered by a small number of cliques; this will be essential for our transformation. Take the family of vertex subsets $\mathcal{G}$ defined as follows: $\mathcal{G} := E(G[Z]) \cup \{V(G) \setminus Z\} \cup \{N_G[v] \setminus (Z \setminus \{v\}) \mid v \in V(G) \setminus Z\}$. The endpoints of an edge trivially form a two-vertex clique, the problem definition ensures that $G - Z$ is a clique, and each vertex together with its neighborhood in the clique $G - Z$ also forms a clique: hence each subset induces a clique in $G$. Since each edge of $G$ is either contained in $G[Z]$, contained in $G - Z$, or connects a vertex in $Z$ to a vertex outside $Z$, the family of cliques covers all edges. It is easy to see that $|\mathcal{G}| \leq \binom{X}{2} + 1 + |X|$, and hence we can cover the edges in the input instance by a number of cliques which is polynomial in the parameter. The clique cover $\mathcal{G}$ can be computed in polynomial time. We use it to construct an output instance as follows.

Initialize $G'$ as a copy of $G$. For each clique $C \in \mathcal{G}$, add a vertex $v_C$ to $G'$ with $N_{G'}(v_C) := C$. Let $Z'$ be the union of $Z$ and all the vertices $v_C$ added in this way. We find that $|Z'| = |Z| + |\mathcal{G}|$ and therefore $|Z'|$ is bounded polynomially in the input parameter $|Z|$; let $k' := |Z'|$. As $G' - Z' = G - Z$, removal of the set $Z'$ from $G'$ results in a clique. We claim that when interpreting the tuple $(G', Z', \ell, k')$ as an instance of FVS or OCT parameterized by a clique deletion set, the answer to the instance is the same as for the input instance $(G, Z, \ell, k)$.

Suppose that $(G, Z, \ell, k)$ is a YES-instance of VERTEX COVER PARAMETERIZED BY CLIQUE DELETION SET, and let $S \subseteq V(G)$ be a vertex cover of $G$ of size at most $\ell$. By the definition of a vertex cover, $G - S$ is an independent set. Since every subset $C \in \mathcal{G}$ is a clique in $G'$, it follows that $|C \setminus S| \leq 1$ for all $C \in \mathcal{G}$. Observe that we can obtain $G' - S$ from $G - S$ by introducing the vertices $v_C$ for all $C \in \mathcal{G}$. Starting from the graph $G - S$ without edges, and using the fact that $|C \setminus S| \leq 1$ for all $C \in G'$ it is easy to see that adding the vertices $v_C$ to obtain $G' - S$ comes down to repeatedly adding vertices of degree at most one to an acyclic graph. As such a process does not create cycles, $G' - S$ is a forest. Therefore $S$ is both an FVS and an OCT for $G'$, of size at most $\ell$, which proves that $(G', Z', \ell, k')$ a YES-instance for the two considered problems.

For the reverse direction, suppose that there is a set $S' \subseteq V(G')$ of size at most $\ell$ whose removal from $G'$ leaves a graph without odd cycles; if $(G', Z', \ell, k')$ is a YES-instance of FVS or OCT then such a set exists. We first show that without loss of generality, we may assume $v_C \notin S'$ for all $C \in \mathcal{G}$. To this end, assume that there is some $C^* \in \mathcal{G}$ such that $v_{C^*} \in S'$. If $C^* \setminus S' = \emptyset$, then $S' \setminus \{v_C^*\}$ is also an OCT for $G'$ as all neighbors of $v_{C^*}$ are contained in the deletion set. If, on the other hand, there exists some $u \in C^* \setminus S'$, then we claim $S^* := (S' \setminus \{u\}) \cup \{v_{C^*}\}$ is also an OCT in $G'$. Indeed, by construction of $G'$ we have $N_{G'}[v_{C^*}] \subseteq N_{G'}[u]$ and therefore we can substitute $u$ for $v_{C^*}$ in any odd cycle in $G' - S^*$ to obtain an odd cycle in $G' - S'$. As the latter graph does not contain odd cycles, it follows $S^*$ is an OCT in $G'$. Since the replacement process can be performed independently for each $C \in \mathcal{G}$ we conclude that if $(G', Z', \ell, k')$ is a YES-instance, then $G'$ has an OCT $S'$ of size at most $\ell$ that does not contain any vertex $v_C$ for $c \in \mathcal{G}$. Now observe that such a set $S'$ forms a vertex cover of graph $G$: if $G - S$ contains an edge $\{x, y\}$, then the edge clique cover $\mathcal{G}$ contains a clique $C_{x,y} \supseteq \{x, y\}$. As $v_{C_{x,y}} \notin S'$ by construction, this implies $G' - S'$ contains a triangle on $x, y$ and $v_{C_{x,y}}$; a contradiction. This establishes the correctness of the transformation. Since it can be carried out in polynomial time, and the resulting parameter $|Z'|$ is bounded polynomially in the input parameter $|Z|$, the theorem follows. □

### 4.4.2 Weighted problems parameterized by vertex cover

In this section we consider the kernelization complexity of weighted versions of FVS and OCT, parameterized by the cardinality of a given vertex cover. Our results extend the line of research initiated by a subset of the authors, who showed that WEIGHTED VERTEX COVER does not admit a polynomial kernel under such a parameterization, unless NP $\subseteq$ coNP/poly [36]. The cross-composition we present unifies a construction from a preliminary version of this work [6] for FVS, with a construction for OCT which appeared in an extended abstract by a subset of the authors [38]. Let us formally define the problems under consideration.

> WEIGHTED FEEDBACK VERTEX SET PARAMETERIZED BY VERTEX COVER
> **Instance:** A graph $G$, a vertex cover $Z \subseteq V(G)$, a weight function $w \colon \mathbb{N} \to \mathbb{N}$, and a positive integer $\ell$.
> **Parameter:** The size $k := |Z|$ of the vertex cover.
> **Question:** Is there a set $S \subseteq V(G)$ that intersects all cycles in $G$ and satisfies $\sum_{v \in S} w(v) \leq \ell$?

The related problem WEIGHTED ODD CYCLE TRANSVERSAL PARAMETERIZED BY VERTEX COVER is defined in the natural way. The parameterized problems are easily seen to be fixed-parameter tractable by noting that the treewidth of the input graphs is bounded by $k$, which allows for a solution by standard dynamic programming techniques [9]. We start by presenting a construction that will be the key to merging multiple input instances.

**Definition 12.** For a graph $G$ define its inflation $\phi(G)$ by the following steps:

- Obtain graph $G'$ from $G$ by completing every edge into a triangle with a new vertex: for each edge $e = \{u, v\} \in E(G)$, add a new vertex $v_e$ with $N_{G'}(v_e) := \{u, v\}$ to $G'$.

- Obtain graph $\phi(G)$ from $G'$ by replacing every edge $\{u, v\} \in E(G')$ by a path on two new vertices: remove edge $\{u, v\}$ while adding vertices $p_1$ and $p_2$ with the edges $\{u, p_1\}, \{p_1, p_2\}$ and $\{p_2, v\}$.

Note that by this definition, the vertex set of $\phi(G)$ is a superset of $V(G)$; hence we may take a subset $S \subseteq V(G)$ and consider the effect of removing $S$ from $\phi(G)$. As an example of the inflation operation, note that the inflation $\phi(K_2)$ of a single edge yields a simple cycle on nine vertices.

**Observation 1.** For any graph $G$ and its inflation $\phi(G)$, the following holds:

- The graph $\phi(G) - V(G)$ is the disjoint union of $3|E(G)|$ copies of $P_2$.

- The set $V(G)$ is independent in $\phi(G)$.

There is a correspondence between vertex covers of $G$, and sets that form an FVS or OCT in the inflation $\phi(G)$. This correspondence will enable us to cross-compose instances of VERTEX COVER into the cycle covering problems.

**Lemma 3.** *For any graph $G$ and integer $\ell$, the following conditions are equivalent.*

1. *The graph $G$ has a vertex cover of size at most $\ell$.*

2. *The graph $\phi(G)$ has a feedback vertex set of size at most $\ell$.*

3. *The graph $\phi(G)$ has an odd cycle transversal of size at most $\ell$.*

*Proof.* (1⇒2) Suppose that $S \subseteq V(G)$ is a vertex cover of $G$ of size at most $\ell$, and consider an arbitrary simple cycle $C$ in $\phi(G)$: we will show that $S$ contains a vertex on $C$. Let $G'$ be the intermediate graph obtained from $G$ that is used in the construction of $\phi(G)$ in Definition 12. As all edges were replaced by $P_2$'s when going from $G'$ to $\phi(G)$, it follows that $C$ can be transformed into a cycle $C'$ in $G'$ by replacing such $P_2$'s by direct edges. Hence $C$ corresponds to a simple cycle $C'$ in $G'$ such that $V(C') \subseteq V(C)$.

- If $C'$ contains a vertex $v_e$ that was introduced to complete an edge $e = \{u, v\}$ into a triangle in the first step of Definition 12, then $v_e$ only has $u$ and $v$ as neighbors and therefore $C'$ contains $u$ and $v$. This pair is connected by the edge $e$ in $G$. As $S$ is a vertex cover of $G$ it contains one of the vertices $u$ and $v$, and therefore intersects the cycle $C'$. Since $V(C') \subseteq V(C)$ the set $S$ intersects $C$ as well.

- If $C'$ does not contain any vertices $v_e$ used to complete edges into triangles, then $V(C') \subseteq V(G)$ and hence the cycle $C'$ also exists in $G$. Now the vertex cover $S$ of $G$ contains at least one vertex of $C'$, which implies that it intersects $C$.

Hence all cycles in $\phi(G)$ are intersected by $S$, which shows that $S$ is an FVS of size at most $\ell$ in $\phi(G)$.

(2⇒3) An FVS of size at most $\ell$ in $\phi(G)$ is also an OCT: the former covers all cycles, and the latter only has to cover odd cycles.

(3⇒1) Suppose $S \subseteq V(\phi(G))$ is an OCT of $\phi(G)$, of size at most $\ell$. We first show that we may assume $S \subseteq V(G)$. So assume $S$ contains a vertex $v \notin V(G)$, i.e., a vertex which was introduced by the inflation operation to subdivide an edge or complete an edge into a triangle. From Definition 12 it follows that $v$ has degree two in $\phi(G)$, and that by following a path of consecutive degree-two vertices in $\phi(G)$ we arrive at a vertex $u$ in $V(G) \cap V(\phi(G))$. As all cycles through $v$ contain the vertex $u$, any cycle intersected by $v$ is also intersected by $u$, and therefore $(S \setminus \{v\}) \cup \{u\}$ is also an OCT of $\phi(G)$. By repeating this step we obtain a set $S \subseteq V(G)$ of size at most $\ell$ which intersects the odd cycles of $\phi(G)$. We claim that this $S$ is a vertex cover in $G$.

Assume for a contradiction that there is an edge $e = \{u, v\}$ in $G$ for which no endpoint is contained in $S$. The graph $G'$ used in the construction of $\phi(G)$ has a vertex $v_e$ adjacent to $u$ and $v$, and hence contains a triangle $C' := (v, v_e, u)$. When creating $\phi(G)$ from $G'$, each edge on this triangle is subdivided by two new vertices resulting in an odd cycle $C$ in $\phi(G)$ on $3 + 6 = 9$ vertices. Hence $\phi(G)$ contains an odd cycle $C$ such that $V(C) \cap V(G) = \{u, v\}$. As $S$ does not contain $u$ nor $v$ by assumption, $\phi(G) - S$ contains an odd cycle which implies that $S$ is not an OCT for $\phi(G)$; a contradiction. □

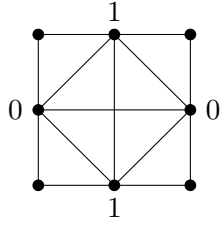

Figure 2: The $K_4$-in-a-box graph $B_{K_4}$ with labeled vertices.

Before presenting the cross-composition we also introduce the following gadget, which will be used as a bit selector.

**Definition 13.** The $K_4$-*in-a-box* graph $B_{K_4}$ (see Figure 2) is the graph obtained from a clique on four vertices $\{a, b, c, d\}$ by adding a new degree-2 vertex $v$ for each pair $\{a,b\}, \{b,c\}, \{c,d\}, \{d,a\}$ such that $v$ is adjacent to both vertices of the pair. The vertices $\{a, c\}$ are the 0-labeled terminals of the graph, and the vertices $\{b, d\}$ are the 1-labeled terminals of the graph.

**Observation 2.** Any vertex set that covers all triangles of $B_{K_4}$ has size at least two. Every size-2 vertex set covering all triangles either consists of the 0-labeled terminals or the 1-labeled terminals.

We present a single construction that can serve as a cross-composition of the classical VERTEX COVER problem into both cycle covering problems considered in this section. Recall that an instance of the NP-complete [30, GT1] VERTEX COVER problem is a tuple $(G, \ell)$ and asks whether $G$ has a vertex cover of size at most $\ell$.

**Theorem 10.** WEIGHTED FEEDBACK VERTEX SET *and* WEIGHTED ODD CYCLE TRANSVERSAL *do not admit polynomial kernels when parameterized by vertex cover, unless* $NP \subseteq coNP/poly$.

*Proof.* We prove the theorem by showing that VERTEX COVER cross-composes into both WEIGHTED FEEDBACK VERTEX SET and WEIGHTED ODD CYCLE TRANSVERSAL under the chosen parameterization. The same cross-composition works for both problems. By a suitable choice of polynomial equivalence relation we may assume the input consists of well-formed instances $(G_1, \ell), \ldots, (G_t, \ell)$ of VERTEX COVER that all have the same number of vertices, say $n$, and edges, say $m$, and that have the same target value $\ell$. By the argument given in the proof of Theorem 8 we may assume that $t$ is a power of 2. As a graph on $n$ vertices trivially has a vertex cover of size $n$, we may output a constant-size YES-instance if $\ell = n$ and assume that $\ell < n$ in the remainder of the proof.

We will merge the inflated versions of the input graphs to obtain the desired cross-composition. By Observation 1 it holds for each input $G_i$ that $\phi(G_i) - V(G_i)$ is a disjoint union of $3m$ copies of $P_2$. Label the $P_2$'s in $\phi(G_i) - V(G_i)$ arbitrarily from 1 to $3m$, and assign each $P_2$ a left and right endpoint. Construct a graph $G'$ with weight function $w'$ as follows.

1. Initialize $G'$ as the disjoint union of the graphs $\phi(G_i)$ for $i \in [t]$.

2. For each $j \in [3m]$, consider the $j$-th copy of $P_2$ in the inflated input graph $\phi(G_i) - V(G_i)$. Modify the graph $G'$ by identifying all the left endpoints of the $j$-th $P_2$ into a single vertex. Similarly identify all the right endpoints of the $j$-th $P_2$ into a single vertex. Let $A' \subseteq V(G')$ contain the resulting merged vertices. Observe that as we merged $P_2$'s one by one, the graph $G'[A']$ is a disjoint union of $P_2$'s and therefore $G'[V(G_i) \cup A']$ is isomorphic to $\phi(G_i)$ for all $i \in [t]$.

3. We can represent an instance number in the range $[t]$ using exactly $\log t$ bits since we assumed $t$ is a power of 2. For each bit position $j \in [\log t]$ we create a copy of the graph $B_{K_4}$ described in Definition 13. We label its 0-terminal vertices $\{b_{j,0'}, b_{j,0''}\}$ and the 1-terminal vertices $\{b_{j,1'}, b_{j,1''}\}$. For each instance number $i$ whose $j$-th most significant bit in the binary expansion is a 0, we make all vertices of $V(G_i)$ in $G'$ adjacent to the 0-terminals $\{b_{j,0'}, b_{j,0''}\}$, and for instance numbers whose bit value is 1 we make it adjacent to $\{b_{j,1'}, b_{j,1''}\}$.

4. Assign each vertex corresponding to a copy of $B_{K_4}$ a weight of $t \cdot n$, and give all other vertices weight 1.

This concludes the description of the graph $G'$ and weight function $w'$. Since a valid instance of WEIGHTED FEEDBACK VERTEX SET parameterized by vertex cover also contains a vertex cover set $Z'$, we must supply such a vertex cover as part of the output of the procedure. Observation 1 shows that for each input $G_i$, the set $V(G_i)$ is independent in $\phi(G_i)$. As $G'$ contains the disjoint union of the inflated input graphs, and since merging the $P_2$'s does not introduce edges in the set $\bigcup_{i=1}^{t} V(G_i)$, it follows that a vertex cover $Z'$ of $G'$ can be formed by taking the union of $A'$ and all vertices of each of the $\log t$ copies of $B_{K_4}$. The size of this vertex cover $Z'$ is $|A'| + 8 \log t = 3m + 8 \log t$. The parameter value $k' := |Z'|$ is bounded by a polynomial in $\log t$ plus the size of the largest instance. We set $\ell' := 2(\log t)t \cdot n + (t-1)n + \ell$. It is easy to see that this construction can be carried out in polynomial time. The constructed tuple can serve as an instance of WEIGHTED FVS PARAMETERIZED BY VERTEX COVER or WEIGHTED OCT PARAMETERIZED BY VERTEX COVER; we will prove that this makes no difference for the truth value. We therefore conclude the proof by showing that the following conditions are equivalent:

1. The graph $G'$ has a feedback vertex set of weight at most $\ell'$.

2. The graph $G'$ has an odd cycle transversal of weight at most $\ell'$.

3. There is an input $i^* \in [t]$ such that $G_{i^*}$ has a vertex cover of size at most $\ell$.

Clearly the equivalence of these conditions shows that the construction serves as a cross-composition of VERTEX COVER into both problems, which establishes the kernelization lower bounds by Corollary 2.

($1 \Rightarrow 2$) Any feedback vertex set is also an odd cycle transversal of the same size.

($2 \Rightarrow 3$) Assume that $G'$ has an OCT $S'$ of total weight at most $2(\log t)t \cdot n + (t-1)n + \ell$. The graph $G'$ contains $\log t$ vertex-disjoint copies of the graph $B_{K_4}$. By Observation 2 at least two vertices are needed to cover all the triangles in $B_{K_4}$, hence the transversal $S'$ contains at least two vertices from each copy of $B_{K_4}$. If there is some copy of $B_{K_4}$ from which $S'$ contains more than two vertices, then as each vertex in a copy of $B_{K_4}$ has weight $t \cdot n$ the set $S'$ has weight at least $(2(\log t) + 1)t \cdot n$, which is strictly greater than $\ell'$ as $\ell < n$. Therefore any OCT $S'$ of weight at most $\ell'$ contains exactly two vertices from each copy of $B_{K_4}$. From Observation 2 it then follows that for each copy of $B_{K_4}$ the set $S'$ contains either the 0-terminal vertices or the 1-terminal vertices, as no other pair of vertices can cover all triangles. We now construct the binary representation of an instance number using the contents of $S'$. Let the $j$-th bit of the number be a 1 if set $S'$ contains $\{b_{j,1'}, b_{j,1''}\}$, and a 0 in the case that it contains $\{b_{j,0'}, b_{j,0''}\}$; let $i^*$ denote the instance number in the range $[t]$ that is represented by this bitstring.

Observe that by the choice of $i^*$, for all vertices in $V(G_{i^*})$ all of their neighbors in the $B_{K_4}$ graphs are contained in $S'$. On the other hand, if we consider some instance number $i' \neq i^*$ then there is at least one bit position where the representations of the numbers $i'$ and $i^*$ differ. Let $j$ be such a bit position and assume for the moment that the $j$-th bit of the number $i^*$ is a 1, which implies that the $j$-th bit of $i'$ is a 0 (the other case is symmetric). Then $S'$ contains the terminal vertices $\{b_{j,1'}, b_{j,1''}\}$ but does not contain $\{b_{j,0'}, b_{j,0''}\}$. But then $S'$ must contain all vertices from the set $V(G_{i'})$, for if $S'$ would avoid some vertex $v \in V(G_{i'})$ then the graph $G' - S'$ would contain a triangle on vertices $\{v, b_{j,0'}, b_{j,0''}\}$ which contradicts the assumption that $S'$ is an OCT. This shows that for all instance numbers $i' \neq i^*$ the set $S'$ contains all vertices of $V(G_{i'})$. These vertices together with the two terminal vertices in each copy of $B_{K_4}$ account for $2(\log t)t \cdot n + (t-1)n$ of the weight of $S'$, and therefore the remaining vertices in $S'$ have weight at most $\ell$; in particular $S'$ contains at most $\ell$ vertices from the set $V(G_{i^*}) \cup A'$ since each such vertex has weight 1. Choose $S'_{i^*}$ as $S' \cap (V(G_{i^*}) \cup A')$; the preceding argument shows $|S'_{i^*}| \leq \ell$. Since $S'$ is an OCT for $G'$ it breaks all odd cycles in all induced subgraphs, hence $S'_{i^*}$ is an OCT of $G'[V(G_{i^*}) \cup A']$. During the construction of $G'$ we already observed that the graph $G'[V(G_{i^*}) \cup A']$ is isomorphic to $\phi(G_{i^*})$, which together with the previous observation shows that $\phi(G_{i^*})$ has an OCT of size at most $|S'_{i^*}| \leq \ell$. By Lemma 3 this implies that $G_{i^*}$ has a vertex cover of size at most $\ell$, thereby proving that $(G_{i^*}, \ell)$ is a YES-instance of VERTEX COVER.

(3⇒1) For the reverse direction, assume that one of the input instances, say $(G_{i^*}, \ell)$, is a YES-instance of VERTEX COVER, i.e., that $G_{i^*}$ has a vertex cover of size $\ell$. By Lemma 3 this implies that $\phi(G_{i^*})$ has an FVS of size at most $\ell$. We construct an FVS $S'$ in $G'$ of weight at most $\ell'$.

1. For each bit position $j \in [\log t]$ add $\{b_{j,0'}, b_{j,0''}\}$ to $S'$ if the $j$-th bit of $i^*$ is a 0, and otherwise add $\{b_{j,1'}, b_{j,1''}\}$. This contributes a total weight of $2(\log t)t \cdot n$.

2. Add the set $\bigcup_{i \neq i^*} V(G_i)$ to $S'$ for a total weight addition of $(t-1)n$.

3. By the construction of $G'$, the graph $G'[V(G_{i^*}) \cup A']$ is isomorphic to $\phi(G_{i^*})$, and by assumption this graph has a feedback vertex set of size at most $\ell$. Add an FVS of $G'[V(G_{i^*}) \cup A']$ of size at most $\ell$ to $S'$; as the vertices in this subgraph have a weight of 1 under $w'$, this increases the weight of $S'$ by at most $\ell$.

Summing up the weight increases we find that the resulting set $S'$ has weight at most $\ell'$. To see that $S'$ indeed intersects all cycles in $G'$, observe that by taking two matching terminal vertices for each copy of $B_{K_4}$ we have broken all cycles within the $B_{K_4}$ graphs. For all sets $V(G_{i'})$ with $i' \neq i^*$ we have taken all the $V(G_{i'})$ vertices into $S'$ so $G' - S'$ cannot contain cycles through such sets $V(G_{i'})$. By taking the appropriate terminal vertices in $S'$ we have broken all connections between vertices in $V(G_{i^*})$ and vertices in copies of $B_{K_4}$. Finally there can be no cycles in $G'[V(G_{i^*}) \cup A'] - S'$ since we have added an FVS for this graph to $S'$. Hence $S'$ indeed covers all cycles, and has the desired weight. □

Slight variations of the proof of Theorem 10 can be used to show kernel lower bounds for various other weighted vertex-deletion problems parameterized by vertex cover, such as CHORDAL DELETION and PLANAR DELETION. Observe that the weight function used in the construction is very simple: it uses only two possible weight values, namely 1 and $t \cdot n$, both of which are polynomially bounded in the total input size to the cross-composition algorithm. The theorem shows that when it comes to weighted problems, even graphs with a simple structure (i.e., graphs with a small vertex cover) can encode the logical OR or many input instances.

## 5 Conclusions

We have introduced the cross-composition framework and used it to derive kernelization lower bounds for structural parameterizations of several graph problems. Through its many appli-cations [7, 8, 14, 15, 16, 17, 19, 31, 36, 38], the framework has already proven itself to be a fruitful tool in the study of kernelization complexity. The key to many cross-compositions is choosing a convenient NP-hard source problem. As the examples in Section 4 show, a series of input graphs $(G_1, \ldots, G_t)$ may be amenable to composition if the vertex set of each graph can be partitioned into $X_i \cup Y_i$ such that all graphs $G_i[X_i]$ are isomorphic, and all graphs $G_i[Y_i]$ are isomorphic. In such settings we may merge the graphs by identifying all subgraphs $G_i[X_i]$ into a single canonical copy. Our construction for CHROMATIC NUMBER employs graphs which decompose into an independent set and disjoint triangles, whereas the inflation $\phi(G)$ of a graph $G$, which underlies the construction of Theorem 10, can be decomposed into an independent set and a disjoint union of $P_2$'s. The importance of starting from a convenient source problem when proving kernel lower bounds can also be observed in the recent work of Dell and Marx [20] and Cygan et al. [16].

Two recent papers [39, 40] extend the cross-composition framework by the use of co-nondeterministic (cross-)compositions. The full version of [39] contains the natural variant of defining a co-nondeterministic OR-cross-composition and a formal proof for the fact that it excludes polynomial kernels and compressions when the dependence on $t$ is at most $f(t) = t^{o(1)}$; the same follows immediately regarding polynomial lower bounds when $f(t) = t^{1/d+o(1)}$. Note that a similar result for AND-cross-compositions is highly unlikely; it is an easy exercise to construct a co-nondeterministic AND-cross-composition for any language. Furthermore, Kratsch et al. [40] point out that OR-cross-compositions do not need NP-hard source problems, but that NP-hardness under co-nondeterministic many-one reductions suffices. To see this, note that the consequence of a cross-composition for $L \subseteq \Sigma^*$ into $\mathcal{Q} \subseteq \Sigma^* \times \mathbb{N}$ together with a polynomial kernel or compression for $\mathcal{Q}$ implies $L \in \text{coNP/poly}$. If $L$ is (only) NP-hard under coNP-many-one reductions then this still implies $\text{NP} \subseteq \text{coNP/poly}$. (This of course holds for any lower bound technique that produces the consequence that $L \in \text{coNP/poly}$.)

We hope to see many applications of cross-composition in the future.

**Acknowledgments.** We would like to thank Holger Dell for insightful discussions.